\documentclass[aps,pra,twocolumn,groupedaddress,citeautoscript]{revtex4-1}
\usepackage[toc,page]{appendix} 
\usepackage[usenames,dvipsnames,svgnames,table]{xcolor}
\definecolor{blue}{rgb}{0.000000,0.000000,1.000000}
\usepackage[plainpages=false,pdfpagelabels,bookmarksnumbered,%
colorlinks=true,%
        linkcolor=blue,
linkcolor=blue,
        citecolor=blue,
citecolor=blue,
        filecolor=blue,
filecolor=blue,
        pagecolor=blue,
pagecolor=blue,
        urlcolor=blue,
urlcolor=blue,
pdftex,%
unicode]{hyperref}
\usepackage{thumbpdf}
\usepackage{physics}
\usepackage{textcomp}
\usepackage{longtable}
\usepackage{graphicx}
\usepackage{amsmath}
\usepackage{amssymb}
\usepackage{combelow}
\usepackage{booktabs}
\usepackage{multirow}
\usepackage{hyperref}
\usepackage{txfonts}
\usepackage[export]{adjustbox} 
\usepackage{soul}
\begin{document}

\title{Floquet topological transitions in 2D Su-Schrieffer-Heeger model: 
interplay between time 
reversal symmetry breaking and dimerization}
\author{Adrian Pena}
\affiliation{National Institute of Materials Physics, Atomi{\cb{s}}tilor 405A, 
	077125 
	M\u{a}gurele \textendash~ Ilfov, Romania}
\author{Bogdan Ostahie}
\affiliation{National Institute of Materials Physics, Atomi{\cb{s}}tilor 405A, 
	077125 
	M\u{a}gurele \textendash~ Ilfov, Romania}
\author{Cristian Radu}
\affiliation{National Institute of Materials Physics, Atomi{\cb{s}}tilor 405A, 
	077125 
	M\u{a}gurele \textendash~ Ilfov, Romania}
\date{\today}
\begin{abstract}
\noindent We theoretically study the 2D Su-Schrieffer-Heeger model in the 
context of Floquet topological insulators (FTIs). FTIs are systems which 
undergo topological phase transitions, governed by Chern numbers, as a result 
of time reversal symmetry 
(TRS)
breaking by a time periodic process.
In our 
proposed model, the condition of TRS breaking is achieved by circularly 
polarized light irradiation. We analytically show that TRS breaking is 
forbidden in the absence of second order neighbors hopping. In the absence of 
light irradiation, we identify a symmetry-protected degeneracy and prove the 
appearance of a flat band along a specific direction in the momentum space. 
Furthermore, we employ a novel method to show that the four unit cell atoms, in 
the absence of irradiation, can be interpreted as conserved spin states. With 
the breaking of TRS via light irradiation, these spin states 
are no longer conserved, leading to the emergence of chiral edge states. 
We also show how the interplay between the TRS breaking and dimerization leads 
to a 
complex phase diagram. The validity of our findings is substantiated through 
Chern numbers, spectral properties, localization of chiral edge states and 
simulations of quantum Hall transport. Our model is suitable not only for 
condensed matter (materials), but also for cold gases trapped in optical 
lattices or electric circuits.

\end{abstract}
\maketitle

\section{Introduction}
The study of topological phases represents a highly topical branch of condensed 
matter physics that has attracted interest from researchers across various 
expertise areas. In the early 1980s, when the Quantum Hall Effect 
\cite{klitzing1980,thouless1982} emerged as one of the most intriguing 
phenomena at the atomic scale, much of the scientific community endeavored to 
explain how charged particles could exhibit such exotic behavior 
(quantification of Hall resistance) under a magnetic field. In the years that 
followed, the study of topological phases transitioned from physical, 
insightful models and thought experiments\cite{laughlin1981} to a 
classification based on symmetry properties \cite{ryu2010,prodan}. Once the 
theoretical foundations were established, 
there was a surge in efforts to discover, synthesize, and characterize new 
topological materials. This led to a significant focus in the experimental 
research community on topological insulators (TIs) 
\cite{zhang2005,chen2009,kuroda2010,tanaka2012,wang2013,atala2013,jotzu2014,xu2016}.
 In more recent times, a notable historical landmark occurred with the Nobel 
Prize awarded to David J. Thouless, F. Duncan M. Haldane, and J. Michael 
Kosterlitz for their "theoretical discoveries of topological phase transitions 
and topological phases of matter" \cite{haldane2017}.

Technically speaking, topological insulators (TIs) are materials that behave as 
common insulators in their bulk, while hosting conducting states at their 
surfaces or edges, depending on the dimensionality of the system 
\cite{haldane2017,moore2010}. Given the robustness of the topological states 
against disorder \cite{hasan2010,qi2011}, these special materials are 
considered the building blocks of future technologies 
\cite{hsieh2009,paudel2013,husale2018,tokura2019,schmitt2022,breunig2022}.

From a physical perspective, the Su-Schrieffer-Heeger (SSH) 1D atomic chain 
\cite{su1980,heeger1988}
serves as the most basic model for understanding topological insulators. 
Developed to explain the insulating behavior of polyacetylene, its simplicity 
allows for analytical approaches and serves as a valuable starting point or 
analogy for studying TIs.

Moving to the 2D class, we must mention the emblematic TI, graphene 
\cite{novoselov2004,geim2007,neto2009,novoselov2009}, which was one of the 
first materials predicted to realize a topological phase transition. However, 
among other lattice models such as Lieb \cite{weeks2010,slot2017} and Kagome 
\cite{sales2019}, the 2D SSH lattice structure 
\cite{liu2017,marques2018,obana2019,ma2022,agrawal2023,wwang2023} has made its 
way recently. It consists of a planar generalization of the SSH chain or, in 
other words, it may be seen as a number of reciprocally connected 1D SSH 
chains, giving rise to a square lattice. Beyond the conceptual model, there are 
numerous possibilities for materializing 2D SSH lattices, including condensed 
matter systems \cite{geng2022}, ultra-cold atomic gases trapped in optical 
lattices \cite{goldman2016,fabre2022} or electric circuits 
\cite{liu2019,wang2023}.

In the physics of TIs, the symmetries play the crucial role. The 2D TIs with 
time-reversal symmetry (TRS) are also known as quantum spin 
Hall insulators, and they are protected by the $Z_2$ 
invariant \cite{kane2005}. When the topological phase transition in a system 
is triggered by the breaking of the TRS, it is categorized as a \textit{Chern 
insulator}. Over recent years, various methods have been 
proposed to 
induce such conditions \cite{kou2017}. What makes this type of TI interesting 
to study is the emergence of so-called \textit{chiral edge states} in the 
 energy gap of system. These states are 
strongly confined to the edges and 
conduct electricity without dissipation, thus sustaining the Quantum Hall 
Effect. The hallmark of a topological phase in a Chern insulator is its 
characterization by an invariant called the \textit{Chern number}, from which 
it derives its name. During the transition between phases, this invariant takes 
on integer values of ±1, ±2, ±3, ..., and is directly related to the 
quantization of the Hall resistance. In the absence of a phase transition, when 
the system is in a trivial (non-topological) phase, the Chern 
number remains 
zero. The most 
common method for breaking the TRS in 
such systems 
involves subjecting them to a magnetic field. However, in a groundbreaking 
paper, Haldane introduced an alternative approach based on imaginary hopping 
between second-order neighbors, employing the hexagonal 
(graphene) lattice as the 
foundational model \cite{haldane1988}, and demonstrated the emergence of a 
non-trivial phase.

More recently, graphene has been utilized as a platform, where it has been 
reported that circularly polarized light irradiation breaks TRS, realizing the 
Haldane model and thus inducing a topological 
phase 
\cite{oka2009,karch2010,kitagawa2010,karch2011,delplace2013,kundu2014, 
perez2014,mikami2016,pena2023,pena2024}. Generally speaking, 
materials that 
undergo a topological phase transition as a result of circularly polarized 
light driving are included in the class of Floquet topological insulators 
(FTIs), named after 
the framework used to model their physics 
\cite{lago2015,li2018,wurl2018,junk2020}. 
Since a FTI in a 
topological phase 
has a non-zero Chern number, circularly polarized light triggers a phenomenon 
known as the 'Quantum Anomalous Hall Effect' \cite{chang2016}. It is termed 
anomalous because traditionally, the Hall effect is associated with the 
presence of a perpendicular magnetic field.

In this paper, we investigate the 2D SSH model from the perspective of FTIs. To 
achieve this aim, we introduce an interaction 
model using a tight-binding approach to elucidate the mechanism of topological 
phase transitions. Initially, we show analytically that TRS breaking is 
unreachable in the absence of diagonal hopping between 
lattice atomic sites. Subsequently, we explore the 2D SSH model in the absence 
of light irradiation and identify a symmetry-protected degeneracy that 
consistently maintains one of the three potential band gaps closed. This 
symmetry, incorporating the time-reversal operation, leads to the 
opening of 
the gap as circularly polarized light breaks the TRS. Furthermore, we employ a 
novel method to show that the four unit cell atoms, in the absence of 
irradiation, can be interpreted as conserved spin states throughout the Fourier 
space. However, with TRS broken by light irradiation, these spin states are no 
longer conserved, and this leads to the emergence of chiral edge states. 
Additionally, the system may undergo a phase transition due to 
dimerization, similarly to Peierls transition  
in SSH models \cite{peierls}. In our model, we distinguish two dimerization 
cases, depending on the inter- and intra-cellular hopping parameters.
We study the topological phases 
by comparing 
these two dimerization cases and highlight an 
intriguing interplay. That is, the system is topologically restricted to always 
maintaining one of its band gaps as trivial in one dimerization state, which 
closes inducing a topological phase transition, when passing into the other 
 state. 
 The phase diagram then gets very complex. We 
corroborate our results with an analysis of spectral features such as band gaps 
and the presence of chiral edge states, in 
addition to Chern numbers. We also explore 
the localization of topological states and show that they are confined to the 
boundaries. Finally, we verify the topological characteristics by simulating a 
4-lead quantum Hall device. 

The present paper is organized as follows: In Section {\ref{I}}, we revisit the 
1D SSH model, briefly describing its emergent topological properties, and 
introduce the dimerization concept. In fact, 
Section \ref{I} 
serves as a benchmark for 
our research. In Section \ref{II}, we elaborate the 2D SSH model and formulate 
the circularly polarized light interaction Hamiltonian within the Floquet 
formalism. As well, we analytically argue that the Floquet topological 
transitions are not allowed in the absence of diagonal direction hopping. In 
Section \ref{III}, we present the main results and give more physical insights 
about the Floquet topological phases and 
dimerization, discussing 
also their interplay. 
We support the findings using an ingenious interpretation of the 2D SSH 
lattice, which allows one to approach the unit cell atoms as spin states. In 
Section \ref{IIII}, we discuss the implications of the light helicity reversal. 
In Section \ref{IIIII}, we summarize our work, reiterating the main results 
and present the conclusions.

\section{SSH Model Revisited}\label{I}

\begin{figure}[h!]
	\begin{center}
		\includegraphics[scale=1.]{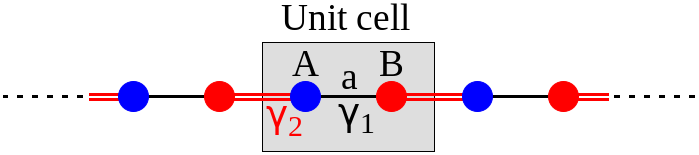}
	\end{center}
	\caption{SSH infinite chain. The unit cell (gray shaded area) contains two 
	atoms indexed by $A$ (blue) and, respectively, $B$ (red). Inside the unit 
	cell, the hopping parameter is $\gamma_1$ and, respectively, between unit 
	cells, $\gamma_2$. The lattice constant is $a$.}\label{fig1}
\end{figure}

The SSH chain represents one of the most important topological models, 
originally introduced to describe the electronic properties of polyacetylene. 
It consists of an infinite bipartite atomic chain, as depicted in Fig. 
\ref{fig1}. The unit cell (gray shaded area) contains two atoms, indexed as $A$ 
(blue) and $B$ (red), respectively, separated from each other by lattice 
constant $a$. The topological properties of the SSH model are conferred by the 
existence of two distinct hopping parameters: $\gamma_1$ in each unit cell and 
$\gamma_2$ between the unit cells. Given its intrinsic simplicity, this system 
explains the fundamental mechanism of topological phase transitions in a 
straightforward manner as an interplay between $\gamma_1$ and $\gamma_2$, as 
will be discussed.

In Fourier space, the Hamiltonian of the system is given by:
\begin{equation}
H(k)=
\begin{pmatrix}
\varepsilon&\gamma_1+\gamma_2 e^{-ika}\\\label{sshham}
\gamma_1+\gamma_2 e^{ika}&\varepsilon
\end{pmatrix},
\end{equation}
where $\varepsilon$ represents the on-site energy.

The topological properties of the SSH chain are governed by the Zak phase 
\cite{zak89}, which represents a Berry-like phase acquired by the wave function 
after one cyclic evolution through the Brillouin zone (BZ), defined as:
\begin{equation}
\varphi_{\text{Zak}}=i\int_{\text{BZ}}\langle
u(k)|\partial_k|u(k)\rangle.\label{zak}
\end{equation}
In the integral (\ref{zak}), which is performed over the entire BZ, 
$|u_n(k)\rangle$ represents the periodic component of the Bloch wave function 
$|\psi(k)\rangle=e^{ikx}|u(k)\rangle$.

\begin{figure}[h!]
	\begin{center}
		\includegraphics[scale=3.]{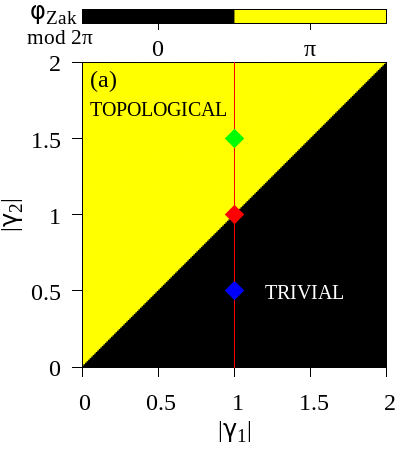}\\
		
		\includegraphics[scale=1.9]{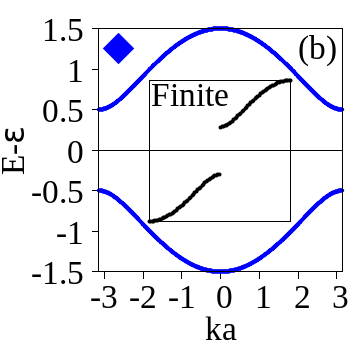}
		\includegraphics[scale=1.9]{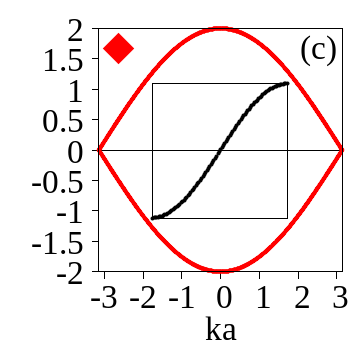}
		\includegraphics[scale=1.9]{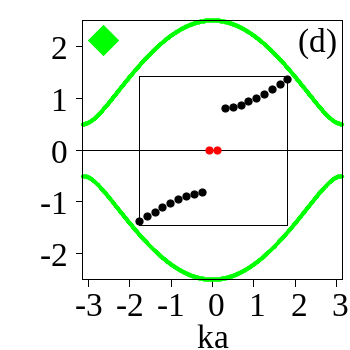}
		\vspace{0.cm}
		
		\includegraphics[scale=1.]{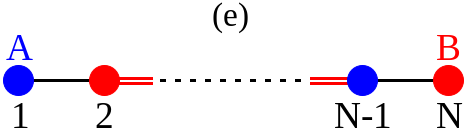}
		\vspace{0.4cm}
		
		\includegraphics[scale=1.9]{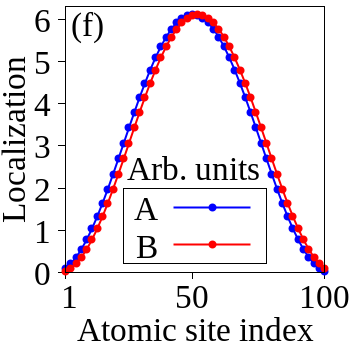}
		\includegraphics[scale=1.9]{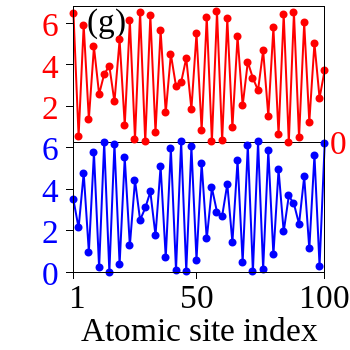}
		\includegraphics[scale=1.9]{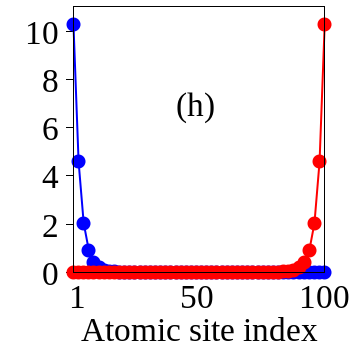}
	\end{center}
	\caption{Topological phases of the SSH chain. (a) Topological phase diagram 
	based on Zak phase: $\varphi_{\text{Zak}}$ mod $2\pi$ vs. $|\gamma_1|$ and 
	$|\gamma_2|$. The system lies in a topological phase whenever $|\gamma_2| > 
	|\gamma_1|$. (b), (c), (d) Energy dispersion for the trivial phase [blue 
	marker within (a), $|\gamma_2|=0.5$], phase transition (red marker, 
	$|\gamma_2|=|\gamma_1|=1$), and topological phase (green marker, 
	$|\gamma_2|=1.5$). The reference level is set at the on-site energy 
	$\varepsilon$. Insets show the eigenenergies for an SSH chain containing 
	$N=100$ atomic sites, as schematized in (e). In the case of the topological 
	phase, two topological states arise inside the gap (red dots). (f), (g), 
	(h) Localization of $A$ and $B$ states corresponding to (b), (c), and (d), 
	respectively.}\label{fig2}
\end{figure}

The inversion symmetry property of the Hamiltonian (\ref{sshham}), 
$\sigma_xH(k)\sigma_x=H(-k)$, leads to the quantization of the Zak phase as 
$\varphi_{\text{Zak}}=0,\pm\pi$ \cite{fu2007,rhim2017,lin2018,li2021}. 
Considering this effect, for $\varphi_{\text{Zak}}=0$, the system lies in a 
trivial (non-topological) phase, and for $\varphi_{\text{Zak}}=\pm\pi$, the 
system undergoes a transition into a topological phase. Based on the Zak phase, 
the topological phase diagram is presented in Fig. \ref{fig2}(a), which depicts 
$\varphi_{\text{Zak}}$ modulo $2\pi$.

Now, we distinguish between two phases: the trivial phase for 
$|\gamma_2|<|\gamma_1|$ (black area) and the topological phase for 
$|\gamma_2|>|\gamma_1|$ (yellow area).
Also, we generally associate the case of $|\gamma_2|<|\gamma_1|$ 
with one dimerization state and, respectively, $|\gamma_2|>|\gamma_1|$ with 
the other possible state.
On the boundary of these two distinct 
phases, the Zak phase is not defined. In what follows, we present the energy 
dispersion for the trivial phase [blue marker within panel (a), 
$|\gamma_2|=0.5$], the topological phase transition (red marker, 
$|\gamma_2|=1$), and the topological phase (green marker, $|\gamma_2|=1.5$) in 
panels (b), (c), and (d), respectively. The energy values are expressed with 
respect to the on-site energy $\varepsilon$ as the reference level. The phase 
transition is fundamentally characterized by the energy band closing at 
$ka=\pm\pi$ when $|\gamma_1|=|\gamma_2|$. In this scenario, the off-diagonal 
coupling terms in the Hamiltonian (\ref{sshham}) vanish, resulting in the band 
closing at the non-bonding combination of $A$ and $B$ orbitals, namely at the 
on-site energy $\varepsilon$. The insets depict the energy eigenvalues for a 
finite SSH chain containing $N=100$ atomic sites, as schematized in Fig. 
\ref{fig2}(e). In the topological phase, two energy values emerge in the gap, 
representing the topological states (depicted as red dots). These states are 
protected by the chiral symmetry $\sigma_zH(k)\sigma_z=-H(k)$. To highlight 
this special property, using Eq. (\ref{ldos}), we illustrate the 
localization within the chain for both 
$A$ and $B$ states in panels (f), (g), and (h), corresponding to the three 
cases presented in panels (b), (c), and (d), respectively. In the trivial 
phase, the atomic states are mainly localized in the middle of the chain. As 
the system evolves towards the topological phase, in the phase transition 
regime, the states become highly delocalized in the bulk of the atomic chain. 
Finally, in the topological phase, edge states emerge, with the $A$ states 
confined to the left edge of the system and the $B$ states localizing at the 
right edge.

The SSH chain topological transition governed by the Zak phase has its origin 
in the so-called dimerization of 1D atomic chains with one electron per atomic 
site and was explained by Rudolf Peierls in 1930. Nowadays, it is called {\it 
Peierls transition} or 
{\it Peierls distortion}. In such systems, the condition for a 
 topological phase transition due to 
dimerization state switching is achieved 
when the intra-cellular hopping parameter equals the inter-cellular one. In 
Fourier space, the transition always takes place at the non-bonding energy 
point.

\section{2D SSH Model. Floquet topological transitions}\label{II}
\subsection{2D SSH model}

\begin{figure}
	\begin{center}
		\includegraphics[scale=1.5]{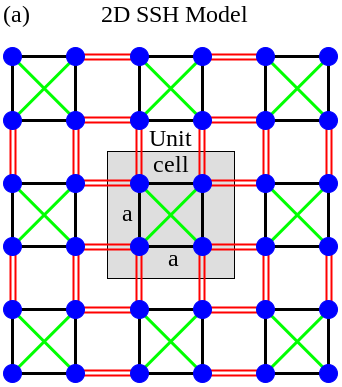}\\
		\includegraphics[scale=1.5]{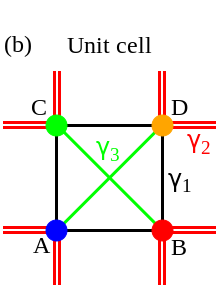}
		\hfil
		\includegraphics[scale=1.6]{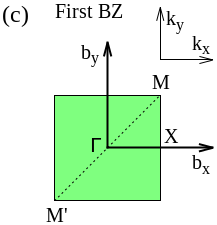}
	\end{center}
	\caption{2D SSH model. (a) The 2D SSH square lattice in real space. The 
		distance between the atoms is denoted by $a$. (b) The unit 
		cell, containing four atoms indexed by $A,B,C$ and $D$. Inside the unit 
		cell, $\gamma_1$ represents the hopping parameter between the first 
		order 
		neighbors (horizontal and vertical) and, respectively, $\gamma_3$, 
		between 
		the second order neighbors (diagonal hopping). (c) First 
		BZ.}\label{fig3}
\end{figure}

In this section, we briefly present the 2D SSH model. We study a square 
lattice system, as presented in Fig. \ref{fig3} (a), with a unit cell (gray 
shaded area)
containing four atoms [see Fig. \ref{fig3} (b)]. The distance between the atoms 
is $a$. In our proposed model, we 
distinguish three hopping parameters: $\gamma_1$ inside each unit cell, 
$\gamma_2$ 
between the unit cells and $\gamma_3$ between the second order neighbor atoms 
(diagonal)
only inside the unit cell. The role of $\gamma_3$ will become clear in the next 
section.

In Fourier space, the system is described by the following Hamiltonian:
\begin{equation}
\begin{aligned}
&H(k_x,k_y)=\\
&\begin{pmatrix}
0&\gamma_1+\gamma_2 e^{-ik_x}&\gamma_1+\gamma_2 e^{-ik_y}&\gamma_3\\
\gamma_1+\gamma_2 e^{ik_x}&0&\gamma_3&\gamma_1+\gamma_2 e^{-ik_y}\\
\gamma_1+\gamma_2 e^{ik_y}&\gamma_3&0&\gamma_1+\gamma_2 e^{-ik_x}\\
\gamma_3&\gamma_1+\gamma_2 e^{ik_y}&\gamma_1+\gamma_2 e^{ik_x}&0
\end{pmatrix}.\label{hamnolight}
\end{aligned}
\end{equation}
Here, for convenience, we have considered a zero on-site energy (diagonal 
elements) and set $a=1$. The First BZ is depicted in Fig. \ref{fig3}(c) 
and the high 
symmetry points are $\Gamma:(0,0)$, $X:(\pi,0)$ and $M:(\pi,\pi)$.

\begin{figure}
	\begin{center}
		\includegraphics[scale=1.6]{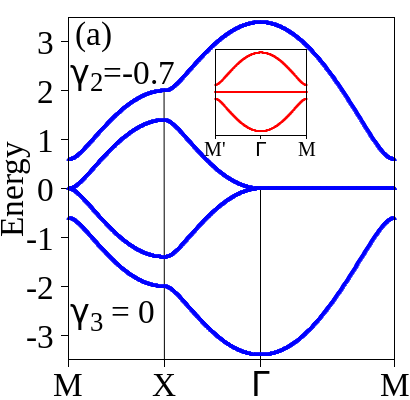}
		\includegraphics[scale=1.6]{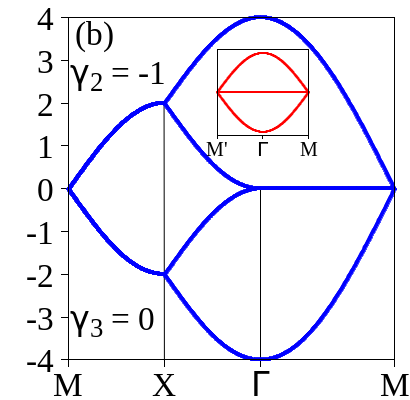}
		\includegraphics[scale=1.6]{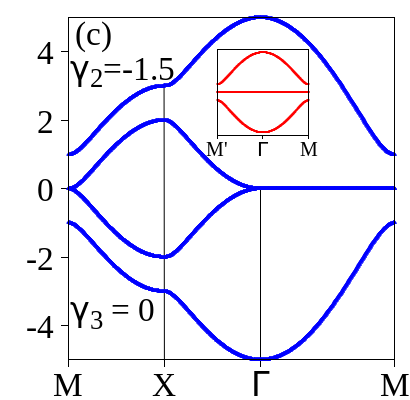}
	\end{center}
	\caption{Energy dispersion of a 2D SSH lattice on $M$-$X$-$\Gamma$-$M$ 
		direction, for $\gamma_1=-1$ and $\gamma_3=0$. The inter-cellular 
		hopping 
		parameters are (a) $\gamma_2=-0.7$ (trivial phase), (b) $\gamma_2=-1$ 
		(topological phase 
		transition) and (c) $\gamma_2=-1.5$ (topological phase). The insets 
		show 
		the dispersion on 
		$M'$-$\Gamma$-$M$ direction. 
	}\label{fig4}
\end{figure}

For the case of $\gamma_3=0$, the 2D SSH model behaves similarly to the 1D SSH 
chain discussed in Section \ref{I}. For instance, considering $\gamma_1=-1$, we 
present in Fig. 
\ref{fig4} the energy dispersion on $M$-$X$-$\Gamma$-$M$ direction in BZ, for 
three 
fundamental cases: trivial phase 
[panel (a), $\gamma_2=-0.7$], phase transition [panel (b), 
$\gamma_2=\gamma_1=-1$] and topological phase [panel (c), $\gamma_2=-1.5$]. The 
insets show the dispersion on $M'$-$\Gamma$-$M$ direction. The gap closing is 
realized by the uppermost and lowermost bands in $M(M')$, while the inner bands 
close permanently on 
$\Gamma$ and $M(M')$. At the phase transition, the gap closing is realized in 
$X$ and $M(M')$. An other important property of the 2D SSH model is the 
occurrence of a completely flat band centered on the 
non-bonding energy level (0 in our present case), on the $M$'-$\Gamma$-$M$ 
direction. This property will be carefully investigated in Section \ref{III}. 
An in-depth presentation may be found in Ref. \cite{obana2019}.

\subsection{TRS breaking: Floquet topological transition}

\begin{figure}
	\begin{center}
		\includegraphics[scale=0.5]{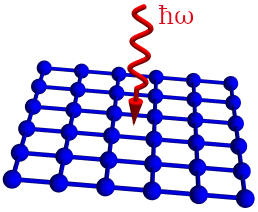}
	\end{center}
	\caption{2D SSH square lattice under circularly polarized light 
		irradiation ($\hbar\omega$). The light is propagating perpendicularly 
		on the lattice plane.}\label{fig5}
\end{figure}

In this section, we introduce an interaction model of 2D SSH system with a 
circularly 
polarized light beam. We consider the light ($\hbar\omega$) propagating 
perpendicularly on 
the lattice plane, as shown in Fig. \ref{fig5}. Our main purpose is to 
elucidate the mechanism of TRS breaking which actually represents the trigger 
of the  
Floquet topological transitions in the studied lattice model.

The physics of the light driven system is modeled using Peierls substitution. 
Generally speaking, within this approach, each hopping parameter contained by 
the Hamiltonian 
(\ref{hamnolight}) is factorized by a time dependent phase 
term $e^{i\theta(t)}$, where
\begin{equation}
\theta(t)=\frac{-e}{h}\int_{\Gamma_{i\rightarrow j}}{\bf A}(t)\cdot{\bf 
ds}.\label{phase}
\end{equation} 
Here, the integral is performed along the path $\Gamma_{i\rightarrow j}$ which 
connects the neighbor atoms $i$ and $j$,
${\bf A}(t)$ represents the light vector potential and ${\bf ds}$ parameterizes 
the integration path. The constant factors $e$ and $h$ represent the 
elementary charge and, respectively, the Planck constant. 

Going further, we 
explicit Eq. (\ref{phase})
considering the following plane wave vector potential:
\begin{equation}
{\bf A}(t)=A_0\left[\cos(\omega t){\bf e}_x+\Lambda\sin(\omega t){\bf
e}_y\right],
\end{equation}
where $A_0$ is a real constant amplitude, $\omega$ the light frequency, 
$\Lambda$ the helicity quantum number, ${\bf e}_x$ and ${\bf e}_y$ the unit 
vectors along the $x-$ and $y-$axis, respectively. Thus, we distinguish now 
four Peierls phases between the first order neighbors (horizontal and 
vertical hopping):
\begin{subequations}
\begin{gather}
\theta_L(t)=\frac{eaA_0}{h}\cos(\omega t);\label{l}\\
\theta_R(t)=\frac{-eaA_0}{h}\cos(\omega t);\label{r}\\
\theta_D(t)=\Lambda\frac{eaA_0}{h}\sin(\omega t);\label{d}\\
\theta_U(t)=\Lambda\frac{-eaA_0}{h}\sin(\omega t)\label{u}.
\end{gather}

Eqs. (\ref{l})$-$(\ref{u}) represent the phases acquired by the electron wave 
function corresponding to a left, right, down and, respectively, up hopping on 
a first order neighbor atomic site. 
\end{subequations}
For the case of second order neighbors hopping (on diagonally placed atomic 
sites), we introduce the following terms: 
\begin{subequations}
	\begin{gather}
	\theta_{LD}(t)=\frac{eaA_0}{h}\left[\cos(\omega t)+\Lambda\sin(\omega 
	t)\right];\label{ld}\\
	\theta_{RD}(t)=\frac{eaA_0}{h}\left[-\cos(\omega t)+\Lambda\sin(\omega 
	t)\right];\label{rd}\\
	\theta_{LU}(t)=\frac{eaA_0}{h}\left[\cos(\omega t)-\Lambda\sin(\omega 
	t)\right];\label{lu}\\
	\theta_{RU}(t)=\frac{eaA_0}{h}\left[-\cos(\omega t)-\Lambda\sin(\omega 
	t)\right]\label{ru}.
	\end{gather}
\end{subequations}
Analogously, Eqs. (\ref{ld})$-$(\ref{ru}) represent the Peierls phases 
corresponding to a left-down, right-down, left-up and, respectively, right-up 
second order neighbors hopping.

Now, considering the matrix representation (\ref{hamnolight}) and the Peierls 
phases (\ref{l})$-$(\ref{u}) and (\ref{ld})$-$(\ref{ru}), we formulate the 
light driven 2D SSH time dependent Hamiltonian:
\begin{widetext}
	\begin{equation}
	H(k_x,k_y;t)=
	\begin{pmatrix}
	0&\gamma_1e^{i\theta_L(t)}+\gamma_2
	e^{i\theta_R(t)}e^{ik_x}&\gamma_1e^{i\theta_D(t)}+\gamma_2e^{i\theta_U(t)}e^{ik_y}
	&\gamma_3e^{i\zeta_{LD}(t)}\\
	\gamma_1e^{i\theta_R(t)}+
	\gamma_2e^{i\theta_L(t)}e^{-ik_x}&0&\gamma_3e^{i\zeta_{RD}(t)}
	&\gamma_1e^{i\theta_D(t)}+\gamma_2e^{i\theta_U(t)}e^{ik_y}
	\\
	\gamma_1e^{i\theta_U(t)}+
	\gamma_2e^{i\theta_D(t)}e^{-ik_y}&\gamma_3e^{i\zeta_{LU}(t)}&0&
	\gamma_1e^{i\theta_L(t)}+\gamma_2e^{i\theta_R(t)}e^{ik_x}\\
	\gamma_3e^{i\zeta_{RU}(t)}&\gamma_1e^{i\theta_U(t)}+
	\gamma_2e^{i\theta_D(t)}e^{-ik_y}&
	\gamma_1e^{i\theta_R(t)}+\gamma_2e^{i\theta_L(t)}e^{-ik_x}&0\\
	\end{pmatrix}.\label{ham}
	\end{equation}
\end{widetext}

Since the Hamiltonian (\ref{ham}) is periodic in time with the period 
$T=\frac{2\pi}{\omega}$, $\omega$ being the light frequency, we will make use 
of the Floquet formalism, described in what follows. The starting point is 
represented by the Floquet theorem 
\cite{shirley1965,giovannini2020} which assures that a time 
periodic 
Hamiltonian 
\begin{equation}
H(t+T)=H(t)\label{perh}
\end{equation}
admits eigenfunctions of the form
\begin{equation}
\psi(t)=e^{\frac{-iWt}{\hbar}}\phi(t).\label{psi}
\end{equation}
In Eq. (\ref{psi}), $W$ is so-called {\it quasienergy} and $\phi(t)$, the {\it 
Floquet 
function}, having the same periodicity as the Hamiltonian (\ref{perh}):
\begin{equation}
\phi(t+T)=\phi(t).\label{perpsi}
\end{equation}

Inserting the form (\ref{psi}) into the Schr\" odinger equation 
$H(t)\psi(t)=i\partial_t\psi(t)$, we end with the following eigenvalue equation 
for the {\it Floquet Hamiltonian} $H_F(t)$:
\begin{gather}
H_F(t)\phi(t)=W\phi(t);\label{floqueteq}\\
H_F(t)=H(t)-i\hbar\partial_t\label{hf}.
\end{gather}
Further, we exploit 
the time periodicity properties (\ref{perh}) and (\ref{perpsi}) to introduce 
the following Fourier representations for the Hamiltonian and the Floquet 
function, respectively:  
\begin{gather}
H(t)=\sum_{n=-\infty}^{\infty}e^{-in\omega t}H_n\label{ht1};\\
\phi(t)=\sum_{n=-\infty}^{\infty}e^{-in\omega 
	t}\phi_n.\label{phit}
\end{gather}
The corresponding inverse transformations are:
\begin{gather}
H_n=\frac{1}{T}\int_0^Te^{in\omega t} H(t)dt;\label{hr}\\
\phi_n=\frac{1}{T}\int_0^Te^{in\omega t}\phi(t)dt.\label{phir}
\end{gather}

Finally, using the Eqs. (\ref{floqueteq})$-$(\ref{phir}), we obtain the {\it 
Floquet system of equations}:
\begin{equation}
\begin{aligned}
&\left(\sum_{m=-\infty}^{\infty} 
H_{m-n}+m\hbar\omega\delta_{mn}\right)\phi_m=W\phi_n;\label{flosys}\\
&\quad\quad\quad n=-\infty,...,-1,0,1,...,\infty.
\end{aligned}
\end{equation}
Note that the mathematical maneuver we have made to obtain the system 
(\ref{flosys}) eliminated the time coordinate degree of freedom.
Despite the system (\ref{flosys}) contains an {\it infinite} number of {\it 
coupled} differential equations, we may formulate effective problems, 
truncating the system in a proper manner, as imposed by the energy scales. If 
the photon energy $\hbar\omega$ is large enough compared to the band width, the 
system is well described by the following high 
frequency Hamiltonian \cite{sentef2015,kristinsson2016}, resulted 
from 
(\ref{flosys}):
\begin{gather}
H_{\text{HF}}(k_x,k_y)=H_0(k_x,k_y)+H_{\text{INT}}(k_x,k_y);\label{hamiltonian}\\
H_{\text{INT}}(k_x,k_y)=\frac{1}{\hbar\omega}\left[H_{-1}(k_x,k_y),H_1(k_x,k_y)\right].\label{hfham}
\end{gather} 
The first term $H_0(k_x,k_y)$ within Eq. (\ref{hfham}) represents the 
averaged 
Hamiltonian 
(\ref{ham}) over a period $T$. The second term $H_{\text{INT}}(k_x,k_y)$ 
describes the virtual interaction 
process of 
absorption-emission of one photon.

\begin{widetext}
\noindent  	
Now, let us evaluate $H_n(k_x,k_y)$, considering $\frac{ea}{\hbar}=1$. 
Performing the integrals (\ref{hr}), we are led to the following substitutions:
\begin{subequations}
\begin{gather}
e^{i\theta_L(t)}\rightarrow i^nJ_n(A_0)e^{i\Lambda n\pi};\label{L}\\
e^{i\theta_R(t)}\rightarrow i^nJ_n(A_0);\label{R}\\
e^{i\theta_D(t)}\rightarrow i^nJ_n(A_0)e^{i\Lambda n\pi/2};\label{D}\\
e^{i\theta_U(t)}\rightarrow i^nJ_n(A_0)e^{-i\Lambda n\pi/2};\label{U}\\
e^{i\theta_{LD}(t)}\rightarrow i^nJ_n(\sqrt{2}A_0)e^{-i\Lambda 
n3\pi/4};\label{LD}\\
e^{i\theta_{RD}(t)}\rightarrow i^nJ_n(\sqrt{2}A_0)e^{-i\Lambda 
n\pi/4};\label{RD}\\
e^{i\theta_{LU}(t)}\rightarrow i^nJ_n(\sqrt{2}A_0)e^{i\Lambda 
n3\pi/4};\label{LU}\\
e^{i\theta_{RU}(t)}\rightarrow i^nJ_n(\sqrt{2}A_0)e^{i\Lambda n\pi/4}.\label{RU}
\end{gather}
\end{subequations}
Here, $J_n(x)$ denotes the $n-$th order Bessel function of the first kind .

Taking into account each harmonic order $n=0,\pm1$ and applying the 
substitutions (\ref{L})$-$(\ref{RU}), we find $H_0(k_x,k_y)$ and 
$H_{\text{INT}}(k_x,k_y)$, having the following form:
	\begin{equation}
	H_0(k_x,k_y)=J_0(A_0)\begin{pmatrix}
	0&\gamma_1+\gamma_2 e^{-ik_x}&\gamma_1+\gamma_2 
	e^{-ik_y}&\frac{J_0(\sqrt{2}A_0)}{J_0(A_0)}\gamma_3\\
	\gamma_1+\gamma_2 
	e^{ik_x}&0&\frac{J_0(\sqrt{2}A_0)}{J_0(A_0)}\gamma_3&\gamma_1+\gamma_2 
	e^{-ik_y}\\
	\gamma_1+\gamma_2 
	e^{ik_y}&\frac{J_0(\sqrt{2}A_0)}{J_0(A_0)}\gamma_3&0&\gamma_1+\gamma_2 
	e^{-ik_x}\\
	\frac{J_0(\sqrt{2}A_0)}{J_0(A_0)}\gamma_3&\gamma_1+\gamma_2 
	e^{ik_y}&\gamma_1+\gamma_2 e^{ik_x}&0
	\end{pmatrix};
	\end{equation}
	
	\begin{equation}
	H_{\text{INT}}(k_x,k_y)=i\gamma_3\frac{
		2\sqrt{2}J_1(A_0)J_1\left(\sqrt{2}A_0\right)}{\hbar\omega}
	\begin{pmatrix}
	0 &\gamma_1-\gamma_2 \cos (k_y) & -\gamma_1+\gamma_2 \cos 
	(k_x) & 0 \\
	-\gamma_1+\gamma_2 \cos (k_y) & 0 & 0 & \gamma_1-\gamma_2 \cos (k_x) \\
	\gamma_1-\gamma_2 \cos (k_x) & 0 & 0 &-\gamma_1+\gamma_2 \cos (k_y) \\
	0 &-\gamma_1+\gamma_2 \cos(k_x) & \gamma_1-\gamma_2 \cos (k_y) & 0 \\
	\end{pmatrix}.
	\end{equation}
\end{widetext}
$H_0(k_x,k_y)$ respects the TRS: 
$H_0(k_x,k_y)=H_0^*(-k_x,-k_y)$. On the other hand, the interaction term 
$H_{\text{INT}}(k_x,k_y)\neq H_{\text{INT}}^*(-k_x,-k_y)$  
breaks the TRS, therefore the virtual process of absorption-emission of 
photons is responsible for the topological phase transitions triggered by the 
light irradiation. However, obviously, this is valid only when $\gamma_3\neq0$. 
Hence, in 
our proposed model, {\it the presence of second order neighbors (diagonal) 
	hopping is 
	mandatory 
	for Floquet topological transitions}. Contrastingly, the Hamiltonian 
remains 
time reversal symmetric and the 2D SSH system lies always in a trivial 
phase. In 
the absence of the diagonal hopping, the only effect of light 
irradiation will be the renormalization of the hopping parameters and, 
implicitly, of the spectrum band width. 
\section{Results and Discussion}\label{III}

In this Section, we present the main results and give more physical insights 
within 
the context of TRS breaking and TIs realm. Since in the 
absence of light irradiation, the 2D SSH model is endowed with some critical 
symmetries, we must start the analysis keeping the light off. By 
analogy with the SSH model described in Section \ref{I}, we distinguish two 
dimerization cases: one for $\gamma_2<\gamma_1$ and, respectively, the other 
one for $\gamma_2>\gamma_1$.
Then, we 
investigate the effect of light driving separately for 
$\gamma_2<\gamma_1$ and, thereafter, for  $\gamma_2>\gamma_1$. Our aim is 
to  
highlight the topological properties of the 
circularly polarized light driven 2D SSH model, comparing the 
two above discussed cases.
\subsection{No light}
First, in this Section, before discussing the Floquet topological transitions 
for some 
specific cases,
we introduce the second order neighbors hopping, in the absence of light 
irradiation. In this case, in 
$\Gamma$ and $M$ within the Fourier space, the eigenvalues of Hamiltonian 
(\ref{hamnolight}) read as 
follows:  
\begin{subequations}
	\begin{gather}
	E_1(\Gamma)=E_2(\Gamma)=-\gamma_3;\label{Egamma}\\
	E_3(\Gamma)=2-2\gamma_2+\gamma_3;\label{Egamma3}\\
	E_4(\Gamma)=-2+2\gamma_2+\gamma_3;\label{Egamma4}\\
	E_1(M)=E_2(M)=-\gamma_3;\label{Em}\\
	E_3(M)=-2-2\gamma_2+\gamma_3;\label{Em3}\\
	E_4(M)=2+2\gamma_2+\gamma_3.\label{Em4}
	\end{gather}
\end{subequations}
	
\begin{figure}[h!]
	\begin{center}
		\includegraphics[scale=2.5]{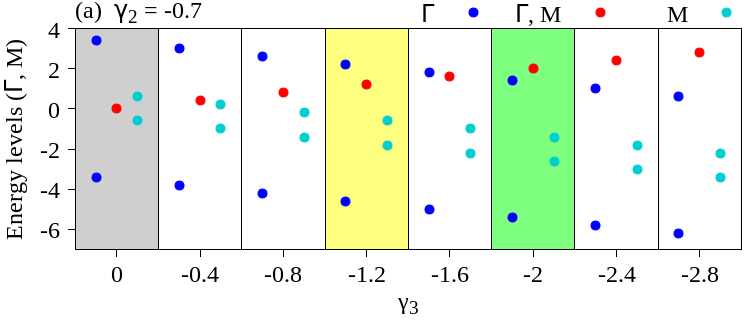}\\
		\includegraphics[scale=1.6]{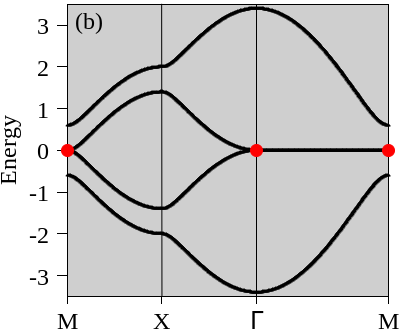}
		\includegraphics[scale=1.6]{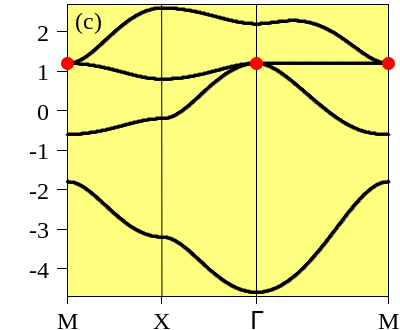}
		\includegraphics[scale=1.6]{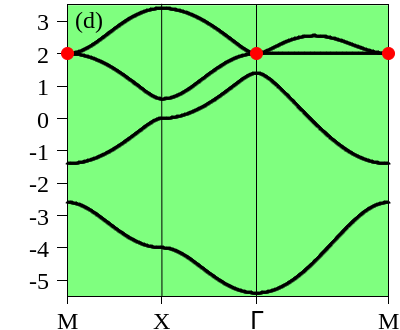}
	\end{center}
	\caption{(a) Energy values in $\Gamma$ and $M$ points vs. $\gamma_3$, with 
	$\gamma_2=-0.7$: blue dots for Eqs. 
	(\ref{Egamma3}) and (\ref{Egamma4}); cyan dots for Eqs. (\ref{Em3}) and 
	(\ref{Em4}); red dots for Eqs. (\ref{Egamma}) and (\ref{Em}). The red dots 
	represent the twofold degenerate energy levels where always two bands 
	crosses. (b), (c), (d): Energy dispersion on $M$-$X$-$\Gamma$-$M$ direction 
	for: $\gamma_3=0$, $\gamma_3=-1.2$ and $\gamma_3=-2$, respectively (as 
	indicated by the background colors). The red dots represent the 
	degeneracy in $\Gamma$ and $M$.}\label{fig6}
\end{figure}

Note that for both $\Gamma$ and $M$, there exist two 
twofold degenerate energy levels with the value $-\gamma_3$, see Eqs. 
(\ref{Egamma}) and (\ref{Em}). Consequently, two bands cross in these two high 
symmetry points at a value of $-\gamma_3$. In Fig. \ref{fig6}(a), we present 
some examples for a 
fixed $\gamma_2=-0.7$ and varying discretely $\gamma_3$. The blue dots 
correspond to Eq. (\ref{Egamma3}) and (\ref{Egamma4}), while the cyan ones, to 
(\ref{Em3}) and (\ref{Em4}), respectively. The red dots represent the twofold 
degenerate energy levels (\ref{Egamma}) and (\ref{Em}). In panels (b), (c) and 
(d), we show the energy dispersion for $\gamma_3=0$, $\gamma_3=-1.2$ and 
$\gamma_3=-2$, respectively. See also the background colors correspondence. For 
each case, 
the bands crossing is marked by the same red dot. 

The twofold degeneracy in $\Gamma$ and $M$ points is a consequence of a 
so-called "hidden symmetry", as discussed in Refs. 
\cite{hou2013,hou2014,hou2018}. The main 
idea is that a (twofold) 
degeneracy is attributed to an 
antiunitary symmetry, expressed by an operator 
$\Upsilon$ with 
its 
square $\Upsilon^2=-1$, satisfying 
$[\Upsilon,H]=0$ in the degeneracy points.

In our problem, in $\Gamma$ and $M$, the normalized eigenvectors of 
Hamiltonian (\ref{hamnolight}) read:
\begin{gather}
|\psi_1\rangle=\frac{1}{\sqrt{2}}\label{psi1}
\begin{pmatrix}
1&0&0&-1
\end{pmatrix}^{\text{T}};\\
|\psi_2\rangle=\frac{1}{\sqrt{2}}\label{psi2}
\begin{pmatrix}
0&1&-1&0
\end{pmatrix}^{\text{T}};\\
|\psi_3\rangle=\frac{1}{2}\label{psi3}
\begin{pmatrix}
1&-1&-1&1
\end{pmatrix}^{\text{T}};\\
|\psi_4\rangle=\frac{-1}{2}\label{psi4}
\begin{pmatrix}
1&1&1&1
\end{pmatrix}^{\text{T}}.
\end{gather} 
The degeneracy corresponds to Eqs. (\ref{psi1}) and 
(\ref{psi2}). 

Further, extending the method presented in Ref. \cite{hou2018}, 
we define the following antiunitary operator:
\begin{equation}
\Upsilon=\left(|\psi_1\rangle\langle\psi_2^*|-|\psi_2\rangle\langle\psi_1^*|+
|\psi_3\rangle\langle\psi_4^*|-|\psi_4\rangle\langle\psi_3^*|\right){\mathcal 
K},\label{ups}
\end{equation}
where ${\mathcal K}$ represents the complex conjugation operator. Substituting 
Eqs. (\ref{psi1})$-$(\ref{psi4}) into Eq. (\ref{ups}), we find the operator in 
question in 
the following form:
\begin{equation}
\Upsilon=
\begin{pmatrix}
0&0&1&0\\
0&0&0&-1\\\label{upsilon}
-1&0&0&0\\
0&1&0&0
\end{pmatrix}
=-i\sigma_y{\mathcal K}\otimes\sigma_z={\mathcal T}\otimes\sigma_z,
\end{equation}
where ${\mathcal T}=-i\sigma_y{\mathcal K}$ denotes the time reversal operator.
Next, the following symmetries satisfied in the degeneracy points may be easily 
verified:
\begin{gather}
[\Upsilon,H(\Gamma)]|\psi_1\rangle=[\Upsilon,H(\Gamma)]|\psi_2\rangle=0;\label{rcom}\\
[\Upsilon,H(M)]|\psi_1\rangle=[\Upsilon,H(M)]|\psi_2\rangle=0.\label{rcom2}
\end{gather} 

Finally, following the proof presented in Ref. \cite{hou2018}, taking into 
account the antiunitarity of $\Upsilon$ and $\Upsilon^2=-1$, one may 
straightforwardly argue that the symmetries (\ref{rcom}) and (\ref{rcom2}) 
protect the twofold degeneracy in $\Gamma$ and $M$.

\begin{figure}
	\begin{center}
		\includegraphics[scale=1.5]{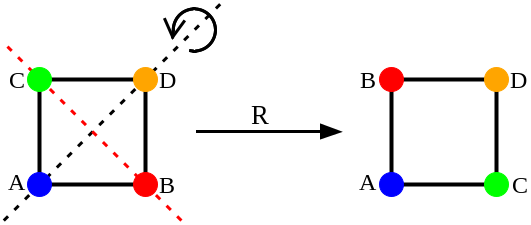}
	\end{center}
	\caption{Reflection of the unit cell with respect to its $A$-$D$ diagonal 
		direction (doted line). The action of ${\mathcal R}$ interchanges the 
		position 
		of $B$ and $C$ atoms.}\label{fig7}
\end{figure}

Having elucidated the underlying physics of the twofold degeneracy 
in $\Gamma$ and $M$, in what follows, we focus our 
attention on the flat 
band arising on 
$\Gamma$-$M(M')$ direction, which is still present regardless of $\gamma_3$ 
magnitude. Also, this spectral property is given by an other 
symmetry, this time being the reflection of the unit cell with respect to its 
diagonal $A$-$D$ or $B$-$C$ axis, as shown in Fig. \ref{fig7}.
The transformation discussed here is expressed by the operator
\begin{equation}
{\mathcal R}=
\begin{pmatrix}
1&0&0&0\\
0&0&1&0\\
0&1&0&0\\
0&0&0&1
\end{pmatrix},\label{RR}
\end{equation}
acting on the state vector $|\psi\rangle=(A,B,C,D)^T$.
One may easily prove that the Hamiltonian (\ref{hamnolight}) obeys the 
following 
symmetry on $\Gamma$-$M(M')$ direction ($k_x=k_y=k$):
\begin{equation}
[{\mathcal R},H(k,k)]=0.\label{rcom3}
\end{equation} 

To explain the occurrence of the flat 
band, we make use of a more subtle interpretation of the system, formulating a 
new effective problem. 
In this respect, taking into account the reflection symmetry discussed above, 
we 
rearrange     
the Hamiltonian (\ref{hamnolight}) in the following block-form:
\begin{gather}
H(k_x,k_y)=
\begin{pmatrix}
h&\Gamma(k_x,k_y)\\
\Gamma^\dag(k_x,k_y)&h
\end{pmatrix};\label{hgamma}\\
h=
\begin{pmatrix}
0&\gamma_3\\
\gamma_3&0
\end{pmatrix};\label{h}\\
\Gamma(k_x,k_y)=
\begin{pmatrix}
\gamma_1+\gamma_2e^{-ik_x}&\gamma_1+\gamma_2e^{-ik_y}\\
\gamma_1+\gamma_2e^{ik_y}&\gamma_1+\gamma_2e^{ik_x}\label{gamma}
\end{pmatrix}.
\end{gather}
Now, starting from the the stationary 
Schr\"odinger eq. $H(k_x,k_y)\psi(k_x,k_y)=E(k_x,k_y)\psi(k_x,k_y)$, one may 
reduce the problem to the following 
effective Hamiltonian:
\begin{gather}
H_{\text{eff}}(k_x,k_y)=
h+\Gamma(k_x,k_y)\left(E\cdot{\boldsymbol 
	1}_{2}-h\right)^{-1}\Gamma^\dag(k_x,k_y).\label{heff}
\end{gather}
In Eq. (\ref{heff}), the symbol ${\boldsymbol 
	1}_{2}$ denotes the identity $2\times2$ matrix. The mathematical maneuver 
	we have implied here is known as {\it decimation method}, see Refs. 
	\cite{pastawski2001,perez2014,mukherjee2022,mukherjee2023}.

As we have reformulated the problem, the Hamiltonian (\ref{heff}) may be also
understood as describing the $B$ and $C$ states, taking into 
consideration the presence of the other two $A$ and $D$ atoms. Now, in order to 
evaluate the system on $\Gamma$-$M(M')$ direction in Fourier space, we
impose the condition $k_x=k_y=k$. After performing all the simplifications, Eq. 
(\ref{heff}) reduces to
\begin{gather}
H_{\text{eff}}(k,k)=
\begin{pmatrix}
\zeta&\zeta+\gamma_3\\
\zeta+\gamma_3&\zeta
\end{pmatrix};\label{heff2}\\
\zeta=2\frac{E(\gamma_2^2+1)+\gamma_3-2\gamma_2(E+\gamma_3)\cos(k)+\gamma_2^2
\gamma_3\cos(2k)}{(E-\gamma_3)(E+\gamma_3)}.\label{zeta}
\end{gather}
In the last step, to obtain the eigenvalues, we must solve the characteristic 
equation
\begin{equation}
\det\Big(H_{\text{eff}}(k,k)-\lambda\cdot{\boldsymbol 
1}_{2}\Big)=0.\label{chareq}
\end{equation}
Straightforwardly, may be verified that Eq. (\ref{chareq}) translates as
\begin{equation}
(\gamma_3+\lambda)f(\lambda;E,k,\gamma_2,\gamma_3)=0,
\end{equation}
where $f$ represents a function of $\lambda$ and depends also on the other 
indicated parameters. It is not necessary to specify here its form. However, it 
is obvious 
that one eigenvalue of the effective Hamiltonian (\ref{heff2}) is 
\begin{equation}
\lambda=-\gamma_3.\label{flat}
\end{equation} 
Since it does not depend on $k_x$ and $k_y$, Eq. (\ref{flat}) expresses exactly 
the flat 
band which arises on $\Gamma$-$M(M')$ direction.

\begin{figure}
	\begin{center}
		\includegraphics[scale=1.6, valign=c]{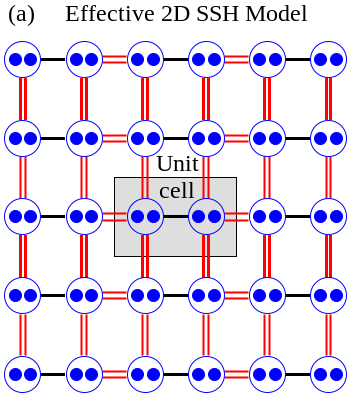}
		\hfil
		\includegraphics[scale=1.4, valign=c]{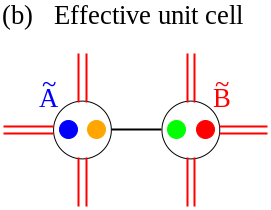}
		\vspace*{\fill}
	\end{center}
	\caption{Effective model. (a) The 2D SSH lattice is reproduced by a 
	bipartite unit cell (gray shaded area). (b) The effective unit cell 
	contains two composite 
	"atoms" indexed by ${\tilde A}$ (formed by the old $A$ and $D$ atoms) and 
	${\tilde B}$ (formed by the old $B$ and $C$).}\label{fig8}
\end{figure}

The decimation method we have involved, allows us to interpret the 2D SSH 
tetraatomic unit cell as an effective {\it bipartite} one, with a non-zero 
and ${\boldsymbol k}$-dependent on-site energy. The effective model thus 
elaborated is depicted in Fig. \ref{fig8}(a). As can be observed here, the 2D 
SSH model is reproduced by the two-component effective unit cell (gray shaded 
area).
The effective unit cell, shown in Fig. 
\ref{fig8}(b), contains two composite "atoms", indexed by $\tilde{A}$ and 
${\tilde B}$, respectively. The "atom" $\tilde{A}$ is formed by the old ones 
$A$ and $D$, while ${\tilde B}$, by old $B$ and $C$.

\begin{figure}
	\begin{center}
		\includegraphics[scale=3]{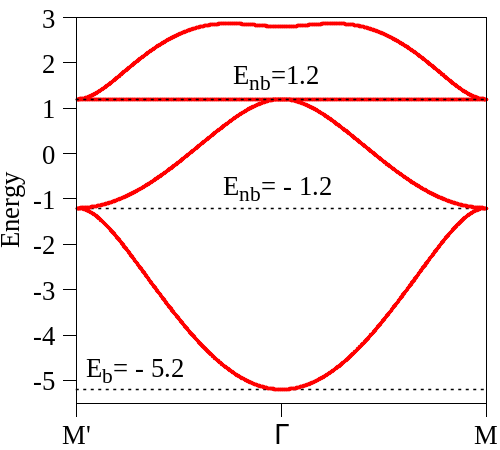}
	\end{center}
	\caption{Energy dispersion on $M'$-$\Gamma$-$M$ direction, for the case of 
	$\gamma_1=\gamma_2$ (topological phase transition condition). The gaps 
	close in $\Gamma$ and $M(M')$ at non-bonding energy $E_\text{nb}=\pm1.2$. 
	The 
	bonding (minimum) 
	energy is $E_\text{b}=-5.2$, in $\Gamma$.}\label{fig9}
\end{figure}

Analyzing Eq. (\ref{zeta}), we find that for 
$\gamma_1=\gamma_2$ in $\Gamma$ and $M$ points of Fourier space [see Fig. 
\ref{fig2}(c)], the effective Hamiltonian (\ref{heff2}) reduces respectively to
\begin{gather}
H_{\text{eff}}(\Gamma)=
\begin{pmatrix}
\frac{8}{E-\gamma_3}&\frac{8}{E-\gamma_3}+\gamma_3\\
\frac{8}{E-\gamma_3}+\gamma_3&\frac{8}{E-\gamma_3}
\end{pmatrix};\\
H_{\text{eff}}(M)=h=
\begin{pmatrix}
0&\gamma_3\\
\gamma_3&0
\end{pmatrix}.
\end{gather}
The corresponding eigenvalues read:
\begin{gather}
E_1(\Gamma)=\gamma_3-4;\quad E_2(\Gamma)=E_3(\Gamma)=-\gamma_3;\quad 
E_4({\Gamma})=\gamma_3+4;\label{eg}\\
E_1(M)=E_2(M)=\gamma_3;\quad E_3(M)=E_4(M)=-\gamma_3.\label{em}
\end{gather}
Since $\gamma_1=\gamma_2$ is the  condition 
for a topological phase transition due to dimerization state switching, we 
interpret Eqs. (\ref{eg}) and (\ref{em}) as follows: $E_\text{b}=\gamma_3-4$ 
represents the 
bonding energy having the minimum value and $E_\text{nb}=\pm\gamma_3$, the 
non-bonding energy. In this case, all the gaps must close in $\Gamma$ and $M$ 
at $E_\text{nb}$ values. See for instance Fig. \ref{fig9} for $\gamma_3=-1.2$, 
with 
$E_\text{b}=-5.2$ and 
$E_\text{nb}=\pm 1.2$.


\subsection{Light irradiation: $\gamma_2<\gamma_1$}

\begin{figure}
	\begin{center}
		\includegraphics[scale=3.5]{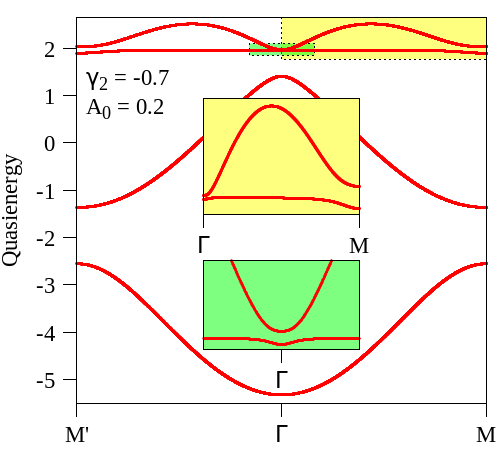}
	\end{center}
	\caption{Energy dispersion on $M'$-$\Gamma$-$M$ direction for the case of 
		$\gamma_2=-0.7$ and $A_0=0.2$. The energy band gap appearance and the 
		flatness breaking of the third band are more clearly revealed in the 
		insets which show a zoomed picture along $\Gamma$-$M$ direction and, 
		respectively, in the vicinity of $\Gamma$.}\label{fig10}
\end{figure}

Next, in this Section, we present the main results concerning the 
Floquet 
topological transitions within the 2D SSH model discussed in Section \ref{II}, 
starting from the following hopping parameters: 
$\gamma_1=-1$ and $\gamma_3=-2$. The photon energy is fixed 
at $\hbar\omega=4$ and the light helicity is $\Lambda=-1$. Throughout of the 
paper, the light vector potential amplitude $A_0$, which gives the driving 
intensity, will be expressed 
in terms of $h/(ea)$ constant. In this configuration, the circularly polarized 
light driven 2D SSH system is described by the Hamiltonian (\ref{hamiltonian}).

From the very beginning, we want to highlight two fundamental effects produced 
by 
the circularly polarized light.
Since the light irradiation breaks the TRS, as argued in Section \ref{II}B, the 
symmetries (\ref{rcom}) and 
(\ref{rcom2}) are no longer satisfied, given that $\Upsilon$ contains the time 
reversal operator ${\mathcal T}$. Hence, the two protected degeneracies will be 
lifted, giving rise to an energy band gap.
Besides the TRS, the light irradiation breaks also the spatial symmetries of 
the system, including also (\ref{rcom3}), and, consequently, the flat 
band occurred in the absence of 
light begins to disperse. These two effects generated by circularly polarized 
light actually represent the precursors of the Floquet topological phase 
transitions. For instance, we show in Fig. \ref{fig10} the quasienergy 
dispersion for the case of $\gamma_2=-0.7$ and $A_0=0.2$. The two effects in 
question are more clearly shown in the insets which present a zoomed picture 
along the $\Gamma$-$M$ direction and, respectively, in the vicinity of $\Gamma$.
In this configuration, having three band gaps, the system is prepared to 
undergo Floquet topological transitions.
 
We begin 
our analysis in terms 
of topological invariants, namely {\it Chern numbers}. Since in our model, the 
circularly 
polarized light breaks the TRS, in the case of 
an infinite system, for each 
energy band there is assigned a possible non-zero Chern number ($C_n$), 
where 
$n$ 
represents the band index, defined as follows: 

\begin{gather}
C_n=\frac{1}{2\pi}\int_{{\text BZ}}{\boldsymbol{\Omega}}_n(\boldsymbol k)\cdot 
d{\boldsymbol k};\label{cn}\\
{\boldsymbol{\Omega}}_n(\boldsymbol k)=\nabla_{\boldsymbol k}\times{\bf 
A({\boldsymbol k})};\label{omega}\\
{\bf A({\boldsymbol k})}=i\langle u_n({\boldsymbol k})|\nabla_{\boldsymbol 
k}|u_n(\boldsymbol 
k)\rangle.\label{A}
\end{gather}
In Eq. (\ref{cn}), the integral is performed over the whole BZ and 
${\boldsymbol{\Omega}}_n(\boldsymbol k)$ defined in Eq. (\ref{omega}), where 
$\nabla_{\boldsymbol k}$ denotes the nabla operator within Fourier space,  
represents the 
Berry curvature. Eq. (\ref A) defines the Berry connection ${\bf 
A}({\boldsymbol k})$ 
with $|u_n({\boldsymbol k})\rangle$ being the periodic component of the Bloch 
wave 
function 
$|\psi({\boldsymbol k})\rangle=e^{i{\boldsymbol k}\cdot{\bf 
r}}|u_n({\boldsymbol k})\rangle$.

\begin{figure}[h!]
	\begin{center}
		\includegraphics[scale=3.5]{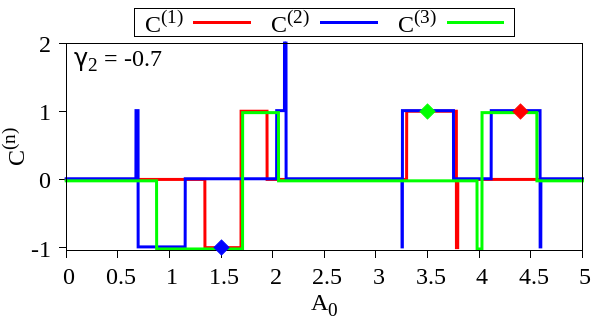}
	\end{center}
	\caption{Topological phase diagram. Summed Chern numbers $C^{(n)}$ for all 
	of
		the three gaps within the quasienergy band structure vs. 
		$A_0$.}\label{fig11}
\end{figure}

Whenever the Fermi level (FL) lies inside an energy gap (assumed to exist due 
to the presence of the light irradiation), one
may describe the 
topological phase of the system summing the Chern numbers for all bands below 
the FL. Thus, the gap in which the FL lies is topologically described by the 
following invariant:
\begin{equation}
C^{(n)}=\sum_{n\in{\text occ}} C_n,
\end{equation}
where the summation in performed over the all $n$ occupied bands (below FL). 

In Fig. \ref{fig11}, we present the topological phase diagram, considering all 
of 
the three quasienergy gaps, with respect to $A_0$. As the $A_0$ value 
continuously varies, we 
observe the 
excitation of different topological phases, governed by an interesting 
restriction. That is, {\it there is impossible to induce topological phases 
for 
the whole three gaps at the same time}. In other words, $C^{(1)}\neq 0$, 
$C^{(2)}\neq 0$ and $C^{(3)}\neq 0$ condition is not achievable simultaneously 
at a fixed 
$A_0$. To explain this topological frustration, we make use again of the 
effective 
model discussed in Section \ref{III}A (Fig. \ref{fig8}). 

First, since the 
effective unit cell contains only two atoms [see Fig. \ref{fig8}(b)], the 
second order neighbors hopping is no longer obvious. On the other hand, within 
the effective model, it becomes a new internal interaction process for "atoms" 
$\tilde{A}$ and $\tilde{B}$. Hence, $A$ and $D$ states become a new internal 
degree of 
freedom for atom $\tilde{A}$ and, analogously, $B$ and $C$ for $\tilde{B}$.  
Moreover, a deeper analysis reveals the following 
symmetry, in the absence of light:
\begin{equation}
[\sigma_x,H_{\text{eff}}(k,k)]=0.
\end{equation}
Therefore, within the atom $\tilde{A}$, $A$ and $D$ act as two well defined 
spin 
states. Obviously, the same is valid for $B$ and $C$ within $\tilde{B}$. Now, 
taking into consideration that generally the spin states are related by 
time 
reversal operation, the TRS breaking due to light irradiation will 
take place at "atomic" level (inside $\tilde{A}$ and $\tilde{B}$), between $A$ 
and $D$ and, respectively, between $B$ and 
$C$.  We anticipate now, that in a topological phase, where chiral edge 
states 
will arise, the mixing between $A$ and $D$; and $B$ and $C$ states is 
topologically 
forbidden. In other words, in a finite configuration, $A$ and $D$ states will 
be 
localized at opposite edges, having also opposite momentum directions and, 
analogously, $B$ and $C$. In Section 
\ref{II}B, we have 
mathematically demonstrated that the presence of second order neighbors hopping 
is 
mandatory for TRS breaking, but with no much physical insight. Having in mind 
the 
interpretation of the 
second order neighbors hopping as an internal interaction process, 
we have now 
a 
deeper physical understanding concerning the TRS breaking in a 2D SSH system 
and its underlying 
mechanism.

Second, going outside the "atoms" $\tilde{A}$ and $\tilde{B}$, which are 
connected only by horizontal and vertical hopping, we do not expect any TRS 
breaking, regardless of light parameters such as intensity, photon energy an so 
on. However, even if there will not occur Floquet topological transitions, the 
system will undergo a topological phase 
transition due to dimerization state switching instead, governed by the 
interplay 
between $\gamma_1$ and $\gamma_2$. Therefore, in the 2D SSH model, there will 
always be reserved an energy gap where the  possible topological 
transition due to dimerization should take 
place. Actually, this is the reason why the simultaneously existence of three 
topological gaps is forbidden. 
\begin{widetext}
	
	\begin{figure}[h!]
		\begin{center}
			\includegraphics[scale=2.2]{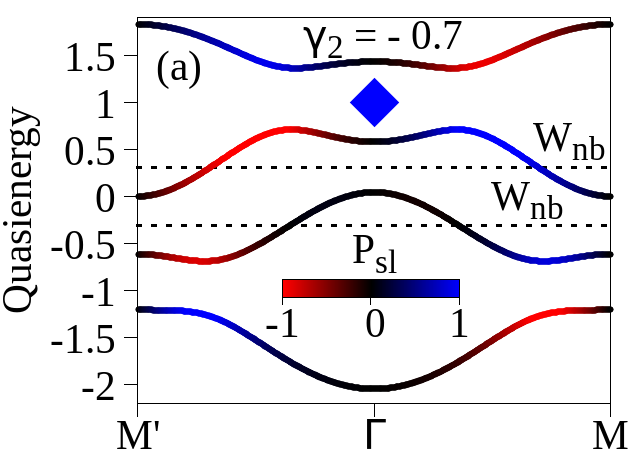}
			\includegraphics[scale=2.2]{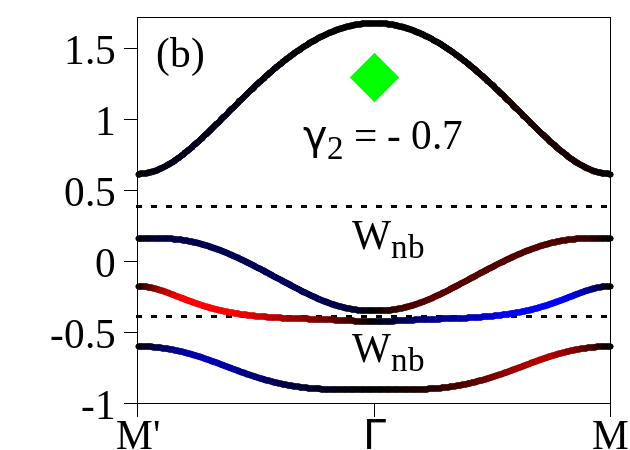}
			\includegraphics[scale=2.2]{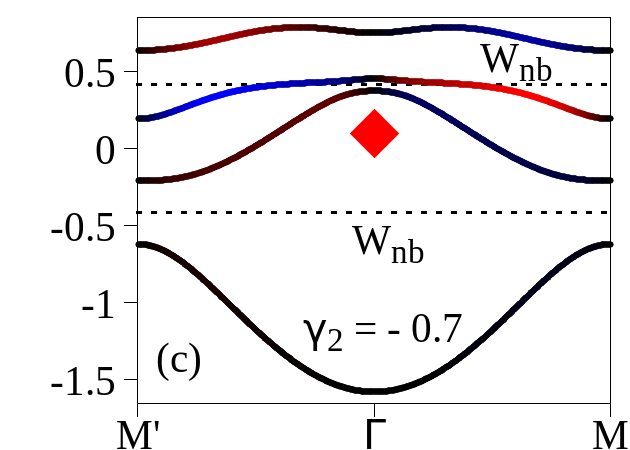}
			\includegraphics[scale=2.2]{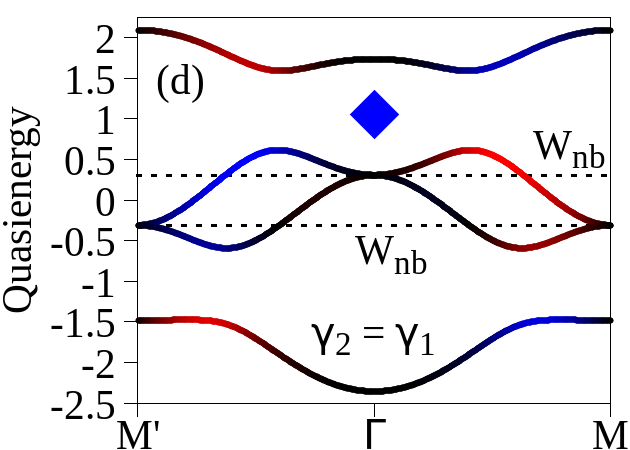}
			\includegraphics[scale=2.2]{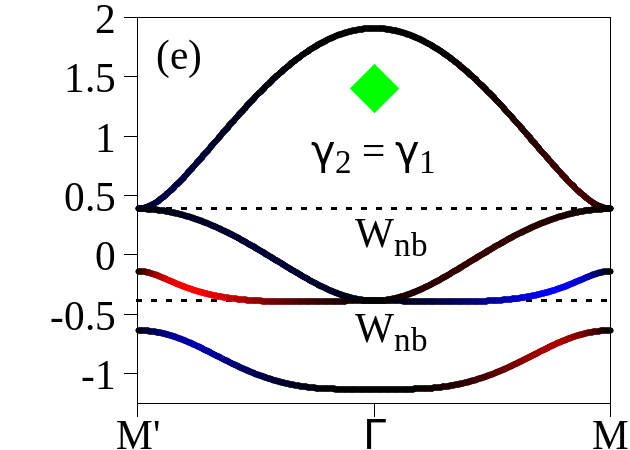}
			\includegraphics[scale=2.2]{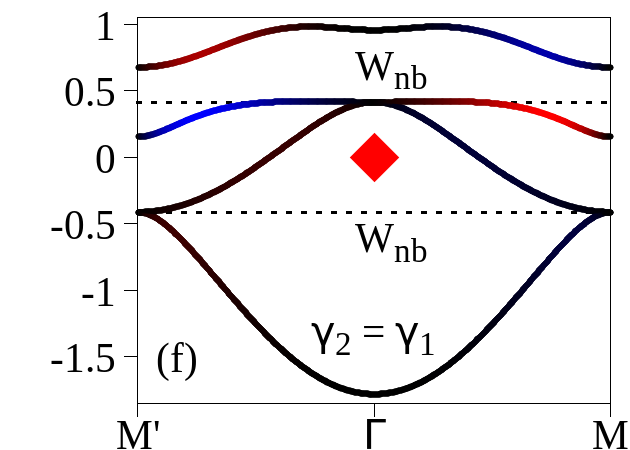}
		\end{center}
		\caption{Quasienergy dispersion on $M'$-$\Gamma$-$M$ direction for 
			different hopping parameters and $A_0$ magnitudes. (a), (b) and (c) 
			show the topological and spectral properties of each phase 
			indicated by 
			blue, green and, respectively, red marker in Fig. 
			\ref{fig11}. The intra-cellular hopping parameter is fixed at 
			$\gamma_2=-0.7$. (d), (e) and (f) illustrate the 
			dimerization state switching scenarios (topological phase 
			transition), 
			achieved for $\gamma_2=\gamma_1$, 
			corresponding to (a), (b) and (c). The horizontal dashed lines 
			represent the two non-bonding quasienergy values. The topological 
			phases 
			characterized by the Chern numbers in Fig. \ref{fig11} reflect in 
			the 
			quasienergy dispersion and the bands polarization. (a) The first 
			and 
			the third gap reveal to be topological, according also to their 
			invariants $C^{(1)}=C^{(3)}=-1$. The  phase 
			transition is 
			realized 
			inside the second (trivial) gap, when $\gamma_2=\gamma_1$, as shown 
			in 
			(d) where the bands which bound the second gap intersect in 
			$\Gamma$ 
			and 
			$M(M')$, at the non-bonding energy value. (b) The first and second 
			gap 
			are topological, having their 
			invariant $C^{(1)}=C^{(2)}=+1$. The third gap (trivial) is reserved 
			for 
			 phase transition, as shown in (e). (c) The second 
			and third gap are 
			topological with $C^{(2)}=C^{(3)}=+1$ and the  phase 
			transition occurs 
			inside the first (trivial) gap, as may be seen in (f).}\label{fig12}
	\end{figure}
	
\end{widetext}
 
The phase transition we are discussing about, 
will arise inside the gap which hosts the non-bonding quasienergy, that is
\begin{equation}
W_\text{nb}=\pm\gamma_3J_0\left(\sqrt{2}A_0\right).\label{wnb}
\end{equation}
Eq. (\ref{wnb}) may be deduced by analogy with the case of light absence. 

Having elucidated this topological property of the 2D SSH model, we 
go further 
and investigate the corresponding spectral properties, associated to three 
topological phases presented in Fig. \ref{fig12} (blue, green and red markers).

First, we consider an infinite system. In order to describe the energy bands 
from a topological point of view, we divide the whole square structure in two 
sublattices, as a preliminary step. Since we have argued that the atoms $A$ and 
$D$ play the role of two different spin states and anticipated that they will 
acquire a well-defined chirality after the TRS breaking, we are lead to group 
$A$ and $B$ atoms in one sublattice, and, respectively, $C$ and 
$D$ in the other. Then, we introduce the sublattice polarization 
defined as
\begin{equation}
P_{\text{sl}}=\frac{\left(|\langle A|A\rangle|^2+|\langle 
	B|B\rangle|^2\right)-\left(|\langle C|C\rangle|^2+|\langle 
	D|D\rangle|^2\right)}{|\langle 
	A|A\rangle|^2+|\langle 
B|B\rangle|^2+|\langle C|C\rangle|^2+|\langle D|D\rangle|^2}.\label{psl}
\end{equation}
$P_\text{sl}\in[-1,1]$ expresses the sublattice population 
weight for each 
quasienergy band, 
at a given point in BZ. For instance, if $P_\text{sl}=1$ at a 
given point 
${\boldsymbol k}$ in BZ, in the $n$-th energy 
band, the contribution is given 
completely by $A$ and $B$ atoms. On the other hand, if 
$P_\text{sl}=-1$, 
there contribute $C$ and $D$. At the middle of these two extreme situations, if 
$P_\text{sl}=0$, the contribution is equal on one hand from $A$ and $B$ and on 
the other from $C$ and $D$. In what follows, we identify the gaps as "first", 
"second" and "third" from the lowest to the highest quasienergy.
The first investigated topological phase 
is for $A_0=1.5$, see blue marker in Fig. \ref{fig11}. The quasienergy 
dispersion is depicted in Fig. \ref{fig12}(a).
The phase diagram 
suggests that in this regime, the first and the third gap lie in a topological 
phase. Specifically for a bipartite topological insulator with TRS, in its 
topological phase, there arises a band polarization (inversion) between the two 
bands which bound the topological gap. Indeed, the sublattice polarization 
function confirms the occurrence of this topological behavior also in our 
studied system, exactly for the first and the third gap. Thus, as the 
bands polarization indicates, in ribbon configuration, we anticipate
the 
crossing of two pairs of chiral bands  in $k=0$ and $k=\pi$, in 
Fourier space. 
 The second gap, trivial, is reserved for the 
topological phase transition.
According to our effective bipartite 
unit cell interpretation (the effective 
model shown in Fig. \ref{fig8}), in 
the actual topological phase, one of the the two non-bonding quasienergy levels 
$W_\text{nb}\approx\pm0.3$, computed from 
Eq. (\ref{wnb}), should be between the first and the third gap (see horizontal 
dashed lines). The bands crossing specific to the topological phase transition 
may be verified 
imposing $\gamma_2=\gamma_1$. In Fig. \ref{fig12}(d), where we show the 
scenario of the  dimerization state switching 
(topological phase transition), one may observe the gap closing in $\Gamma$ 
and $M$, at $W_\text{nb}$.

Next, we analyze the phase indicated by the red marker in Fig. \ref{fig11}, 
where $A_0=3.5$. We expect to find the first and the second gap in a 
topological phase. Fig. \ref{fig12}(b) confirms and, moreover, we see the 
specific 
localization of the non-bonding energy inside the third, trivial, gap. In Fig. 
\ref{fig12}(e), the SSH phase transition is shown. Interestingly, in this 
regime, the two non-bonding energy values does no longer lie in the same gap. 
Finally, in the phase corresponding to the red marker, we would have to find in 
topological phase the second and third gap and $W_\text{nb}$ inside the first, 
trivial, gap. Fig. \ref{fig12}(c) shows this topological configuration. Also, 
the topological phase transition is 
illustrated in Fig. \ref{fig12}(f).

\begin{figure}[h!]
	\begin{center}
		\includegraphics[scale=1.5]{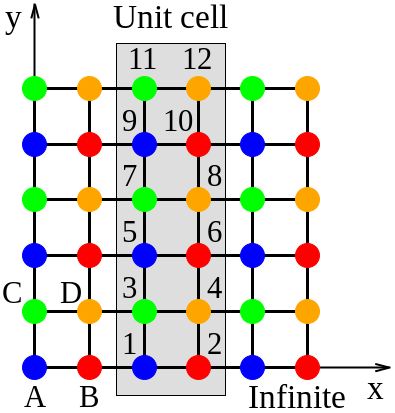}
	\end{center}
	\caption{2D SSH ribbon. The system is kept infinite on $x-$direction 
		(horizontal) and confined on $y-$direction (vertical). The unit cell 
		(gray 
		shaded area) contains the sequence $A,B,C,D,...,D$ atoms indexed by 
		$1,2,3,4,...,12$.}\label{fig13}
\end{figure}

Now, we investigate the topological properties of a 2D SSH ribbon under 
circularly polarized light irradiation, with the
same parameters as above. We 
keep the 
system infinite on $x-$direction (horizontal), thus $k_x$ is still a good 
quantum number and consider the lattice confined on $y-$direction. Since 
the 
periodic boundary condition becomes impossible to apply on $y-$direction, the 
old $k_y$ good 
quantum number is replaced by an atomic index. For instance, in Fig. 
\ref{fig13}, we show a 2D SSH ribbon, where the unit cell (gray shaded area) 
contains the atoms sequence $A,B,C,D,...,D$ indexed by 
$1,2,3,...,12$.

\begin{widetext}
	
	\begin{figure}[h!]
		\begin{center}
			\includegraphics[scale=2.1]{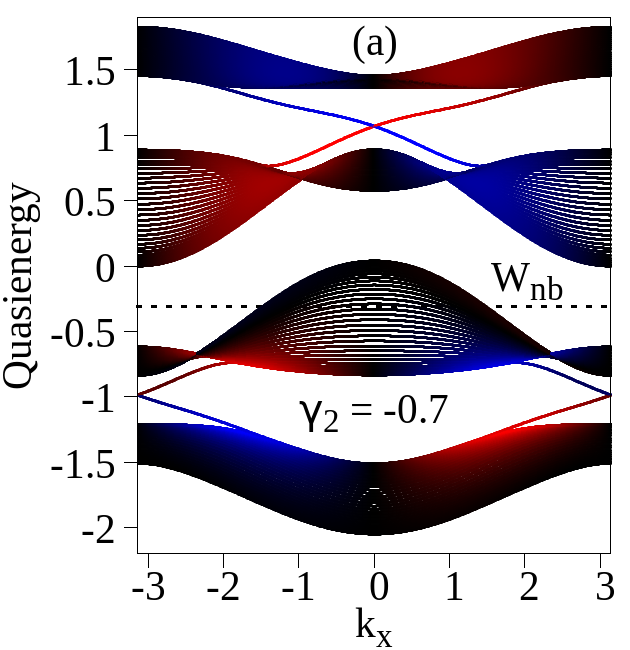}
			\includegraphics[scale=2.1]{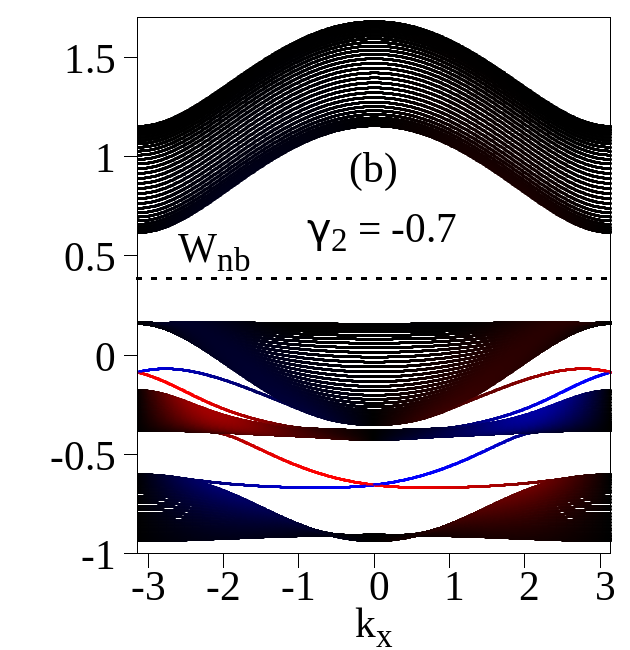}
			\includegraphics[scale=2.1]{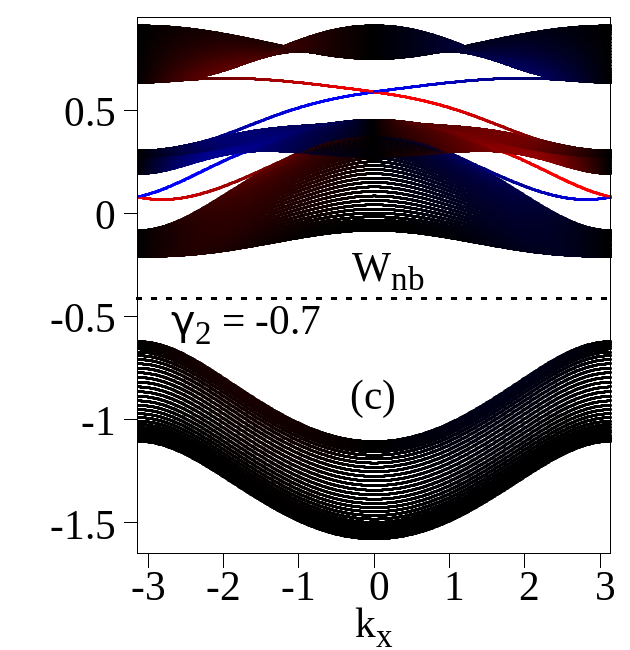}
		\end{center}
		\caption{2D SSH ribbon quasienergy dispersion corresponding to: (a) 
			blue, (b) green and, respectively, (c) red marker in Fig. 
			\ref{fig11}. (a) The first and third gap are topological, according 
			to 
			their invariant $C^{(1)}=C^{(3)}=-1$. Inside each of the two 
			topological 
			gaps, there arise
			 pairs of edge states. The 
			positive momentum band is 
			populated 
			by $C$ and $D$ states, while the negative momentum direction 
			band, 
			by $A$ 
			and $B$. (b) The first and second gap are topological, having the 
			invariants $C^{(1)}=C^{(2)}=+1$. In this case the chirality of the 
			topological states reverses. (c) The second and third gap are 
			topological, 
			with 
			$C^{(2)}=C^{(3)}=+1$. The chirality of the topological states is 
			the 
			same as in (b). The non-bonding energy lies inside the trivial gap 
			for 
			all the three cases.}\label{fig14}
	\end{figure}
	
\end{widetext}

As predicted, in a ribbon, inside the topological gaps, there 
will arise pairs of chiral edge states. In the 
first case (blue marker), the 
chiral edge states occur inside the first and third gap,
which have been identified in a 
topological phase, see Fig. \ref{fig14}(a). According to the value of the 
topological invariants 
$C^{(1)}=C^{(3)}=-1$, inside each gap will arise only one pair.
The topological 
signature of the two quasienergy bands in question resides in their 
polarization. As one may observe, the positive direction momentum band is 
populated by $C$ and $D$ states, while the negative momentum direction band, 
by $A$ and $B$. As well, since both invariants possess the same sign, the 
chirality of the topological states is the same for both gaps. Since besides 
the polarization, an other important topological signature is represented by 
the strong confinement of the topological states at the edges of the system, 
there will arise conduction channels with opposite direction, localized at the 
systems extremities. This property will be investigated a little bit later.
In Fig. \ref{fig14}(b), we show the energy dispersion corresponding to 
the second discussed topological phase (red marker). In this case, the phase is 
characterized by $C^{(1)}=C^{(2)}=+1$. 
Consequently, also in this regime, inside 
each topological gap (first and second) will arise only one pair 
of chiral 
bands. On the other hand, since the sign of the topological invariants is 
reversed, the chirality of the edge states is also reversed. Other way to say, 
at the edges of the system, the localization will not be affected, but, 
instead, 
the momentum direction will be inverted. However, this topological effect is 
obvious inside the topological gap, where one may observe that, unlike in the 
previous case, the positive 
momentum direction chiral band is populated by $A$ and $B$ states, while the 
negative momentum direction band, by $C$ and $D$. 

Finally, the third phase  
regime (red marker), see Fig. \ref{fig14}(c),
characterized by $C^{(2)}=C^{(3)}=+1$, also confirms our expectations,        
 according to the phase diagram and the analysis of the infinite system [Fig. 
 \ref{fig12}(c)]. As well, in the ribbon configuration, the non-bonding energy 
 lies inside the trivial gap. 
 
 For a further description of the Floquet topological phases of 2D SSH model, 
 we 
 intend to extend a little the above discussion. 	
In order to highlight the fundamental characteristics of the found topological 
states in our system, we make use of the local density of states function 
(LDOS), 
defined as
\begin{equation}
\text{LDOS}=\frac{-1}{\pi}\Im[G(i,i,E_F)],\label{ldos}
\end{equation}
where $G(i,i,E_F)$ represents the Green's function of the system Hamiltonian, 
$i$ is the atomic index and $E_F$ the value of FL. 

\begin{figure}[h!]
	\begin{center}
		\includegraphics[scale=2.5]{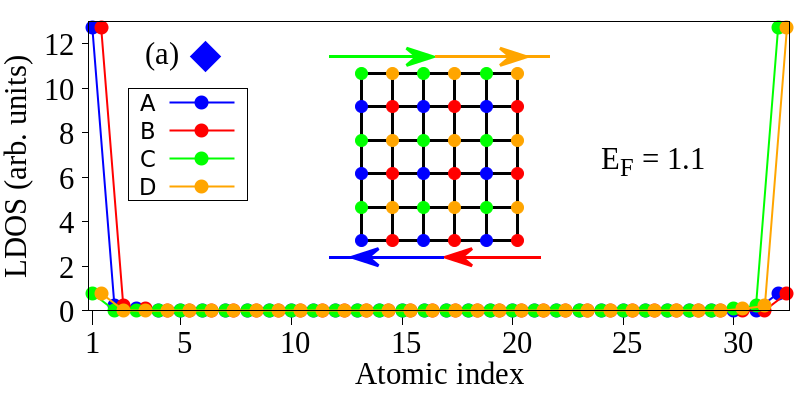}
		
		\vspace{0.3cm}
		\includegraphics[scale=2.3]{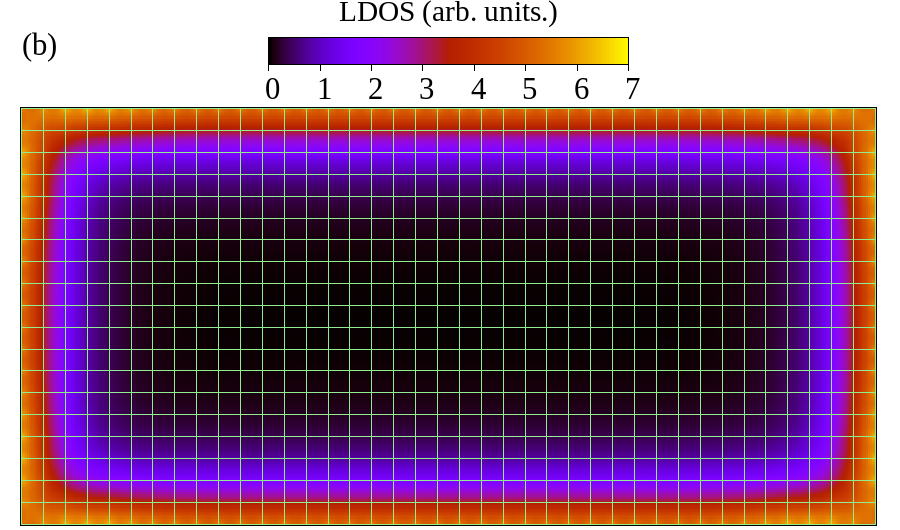}
	\end{center}
	\caption{LDOS for a (a) 2D SSH ribbon and, respectively, (b) for a 2D SSH 
		sheet. The system is in the topological phase indicated by the blue 
		marker 
		in 
		Fig. \ref{fig11}. (a) The FL is considered at $E_F=1.1$, hence inside 
		the 
		third 
		gap of the quasienergy dispersion shown in Fig. 
		\ref{fig14}(a). At the bottom edge of the system, are localized $A$ and 
		$B$ 
		states, while at the top edge, $B$ and $C$, respectively. The inset 
		represents a sketch of the conduction channels 
		formed at the edges of the ribbon. The upper channel transports $B$ 
		and $C$ 
		states in the positive direction, while the lower, $A$ and $B$ states 
		in 
		the negative direction. (b) The system contains $40\times20$ atomic 
		sites.}\label{fig15}
\end{figure}

We chose, for instance, the 
first analyzed phase (blue marker) and consider the FL at $E_F=1.1$. The 
results are depicted in Fig. 
\ref{fig15}(a) for a ribbon with a unit cell containing a number of 32 atoms. 
The distribution of the LDOS values, for each $A,B,C$ and $D$ state, highlights 
the strong confinement of the topological states at the ribbon edges. As well, 
this 
\begin{figure}[h!]
	\begin{center}
		\includegraphics[scale=1.5]{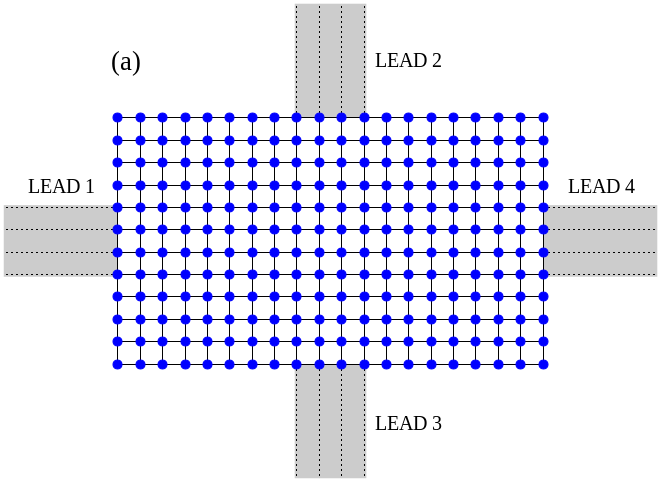}
		
		\vspace{0.2cm}
		
		\includegraphics[scale=2.]{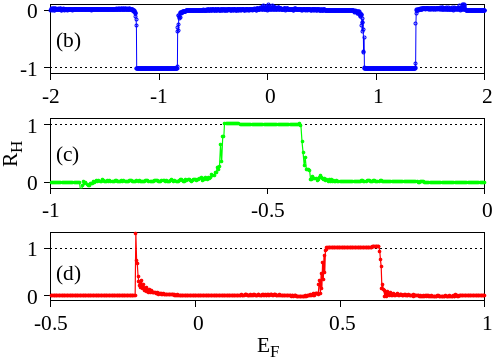}
	\end{center}
	\caption{Quantum Hall transport. (a) Hall device design. The 2D SSH lattice 
		has four leads attached, formed by a number of non-interacting 
		semi-infinite identical
		atomic chains, having the role only to inject and collect 
		charge. (b), (c), (d) Quantum Hall transport simulation using 
		Landauer-B\"uttiker formalism, for the case of blue, green and, 
		respectively, red marker in Fig. \ref{fig11}. Only (b) shows the two 
		intuited $R_H$ 
		plateaus, according to the corresponding Chern numbers of the two 
		topological gaps. In (c) and (d), there is formed only one plateau, 
		given 
		that inside one of the two gaps, the edge states are mixed with the 
		bulk 
		ones.}\label{fig16}
\end{figure}
analysis confirms our prediction that the mixing between $A$ and $D$ 
states 
is topologically forbidden, as a consequence of TRS breaking inside the 
composite "atoms" ${\tilde{A}}$ and $\tilde{B}$ within the effective model 
represented in Fig. \ref{fig8}. The same is valid for $B$ and $C$ states. The 
inset sketches the formation of the conduction channels at the system 
extremities. 
According to Fig. \ref{fig13} and the LDOS, at the bottom of the system, the 
conduction channel will transport $A$ and $B$ states, in the negative direction 
(from right to left). Analogously, at the top, the conduction channel will 
transport $C$ and $D$ states in the positive direction. Moreover, if we cut the 
ribbon on its infinite direction (finite sheet configuration), the chiral edge 
states will confine at the all four edges. In Fig. \ref{fig15}(b), we show the 
LDOS for a sheet formed by $40\times20$ atoms.

Finally, we simulate a Hall transport experiment, based on Landauer-B\"uttiker 
formalism. See Fig. \ref{fig16}(a) for the device design. We consider a 2D SSH 
finite lattice which has four leads attached, whose role is only to inject 
and collect charge. The leads are considered to be formed by a number of 
semi-infinite identical atomic chains which 
do not interact each other. 
For instance, we apply the potentials $V_1$ and $V_2$ on 
lead 1 and, respectively, lead 2 and measure the Hall resistance between lead 
3 
and lead 4. See Appendix A for a brief derivation of the Hall resistance $R_H$.

In Fig. \ref{fig16}, panels (b), (c) and (d), we present $R_H$ in terms of 
$h/e^2$ constant, with respect to 
$E_F$, for the three topological phases analyzed above (blue, green and red 
markers in Fig. \ref{fig11}). The range of $E_F$ is appropriately chosen to 
keep only the relevant results. The most obvious Hall Effect is for the first 
case [panel (b)], where two $R_H$ plateaus are well defined at $R_H=-1$, 
corresponding to the first and the third gap of the quasienergy dispersion 
shown in Fig. \ref{fig14}(a) (topological gaps). Contrastingly, the next two 
cases [panels (c) and 
(d)] are not showing the intuited behavior, according to the Chern numbers in 
Fig. \ref{fig11}. As may be seen, only 
one gap has pure topological states [first gap in (b) and third in (c)], while 
inside the second one, the topological states are mixed with the bulk ones. 
Thus, the Hall Effect is quenched for any $E_F$ value inside those gaps. 
However, the Quantum Hall Effect may be activated even in this situation, by 
inserting lattice defects in order to localize the bulk states, giving rise to 
a so-called {\it Topological Anderson Insulator} 
\cite{li2009,ostahie2018,stutzer2018}.

\subsection{Light irradiation: $\gamma_2>\gamma_1$}
In this Section, we approach the case  of $\gamma_2>\gamma_1$.
We consider that restricting ourselves to phase diagram, ribbon dispersion 
and Hall transport is enough to compare to the previous situation. The 
results presented here are obtained setting $\gamma_2=-1.5$, keeping all the 
other parameters at the same value.

\begin{figure}[h!]
	\begin{center}
		\includegraphics[scale=3.5]{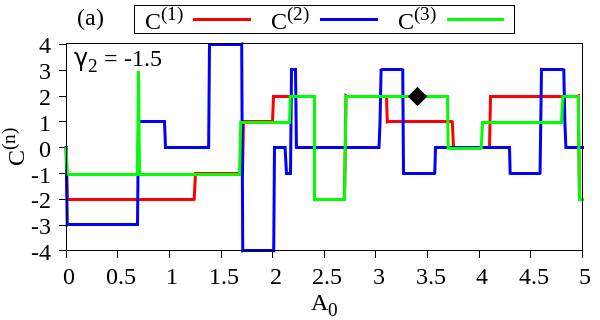}
		
		\vspace{0.2cm}
		
		\includegraphics[scale=2]{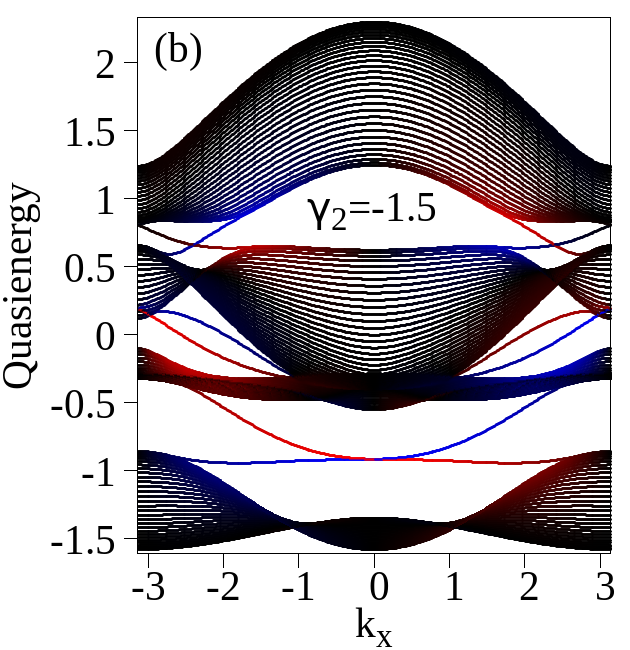}
		
		\vspace{0.3cm}
		
		\includegraphics[scale=4]{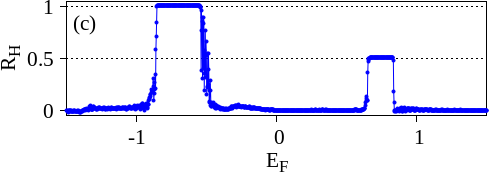}
	\end{center}
	\caption{Topological properties for 
	the case of
	$\gamma_2>\gamma_1$. (a) Topological  
		phase diagram. Here, we have examples where the three invariants are 
		simultaneously non-zero, 
		situation forbidden  for 	
		$\gamma_2<\gamma_1$. (b) Ribbon quasienergy 
		dispersion for the phase indicated by the black marker in (a), 
		$A_0=3.4$. 
		All the three gaps 
		host chiral edge states, even though the second 
		one combines edge with bulk 
		states. (c) Hall transport simulation. The first $R_H=+1$ plateau 
		corresponds to the first gap within (b). The second gap does not have 
		an 
		associated plateau, since there exists a combination of edge with bulk 
		states. The $R_H=+\frac{1}{2}$ plateau is associated with the third gap 
		and 
		it represents the signature of the 
	new topological phase 
			induced as a result of dimerization state switching.}\label{fig17}
\end{figure}

Fig. \ref{fig17}(a) shows the phase diagram, which is obviously more intricate 
than the previous one (Fig. \ref{fig11}). Moreover, we may have several 
examples 
where the three invariants are simultaneously non-zero, which, as discussed, 
for $\gamma_2<\gamma_1$, it is a 
forbidden configuration. 

Fig. 
\ref{fig17}(b) illustrates the quasienergy dispersion for a ribbon, 
corresponding to the 
black marker in Fig. \ref{fig17}(a), namely for $A_0=3.4$. The spectrum is 
quite similar with Fig. \ref{fig14}(b) which is computed for $A_0=3.5$. 
However, in the present case, the third gap is topological, hosting two pairs 
of chiral bands. Comparing Figs. \ref{fig14}(b) and \ref{fig17}(b), we conclude 
that the dimerization state 
switching triggers a new topological phase governed by a Chern number 
$C^{(3)}=+2$. This new phase transition 
occurs inside the third gap [see also Fig. 
\ref{fig12}(e)]. Now, we have simultaneously all the three gaps topological, 
even 
if the second one mixes edge with bulk states. 

 Finally, we simulate the Hall transport. As in the case of Fig. 
 \ref{fig16}(c), the $R_H=+1$ plateau, corresponding to the first gap, is 
 present 
 and very well defined. As $E_F$ rises, reaching the second gap, the Hall 
 conduction is still absent as a consequence of edge and bulk states 
 combination. Specific to our topological configuration, in the Hall resistance 
 diagram, an $R_H=+\frac{1}{2}$ plateau arises, given the existence of the 
 chiral edge states inside the third gap. This plateau 
 represents the 
 signature of the new topological phase 
 induced as a result of dimerization state switching. 
\section{Final remarks: Light Helicity reversal} \label{IIII}

\begin{figure}[h!]
	\begin{center}
		\includegraphics[scale=2]{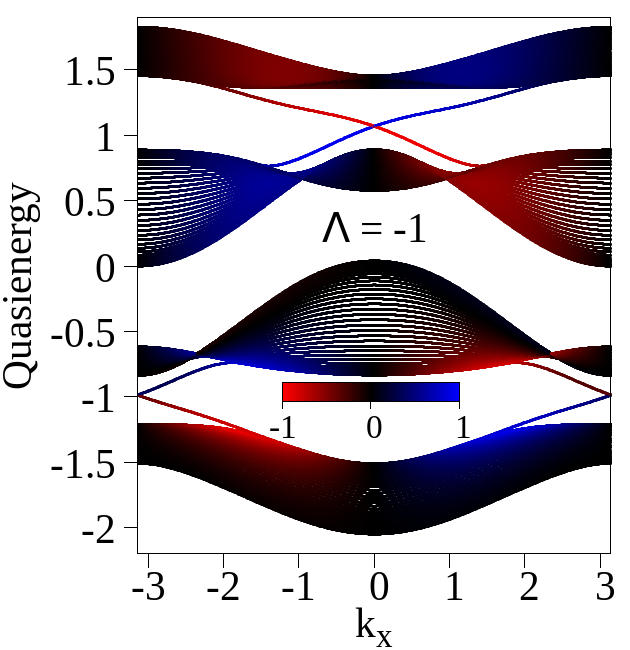}
	\end{center}
	\caption{The counterpart of Fig. \ref{fig14}(a), for $\Lambda=+1$. The 
		particles chirality is reversed, as may be observed from the bands 
		polarization: $A$ and $B$ states will have a positive momentum, while 
		$C$ 
		and 
		$D$, a negative one.}\label{fig18}
\end{figure}

Since the trigger of the Floquet topological phase transitions is the 
circularly 
polarized light, it is worth discussing also its helicity reversal case 
$\Lambda\rightarrow-\Lambda$. The clue resides in Eqs. (\ref{ld})$-$(\ref{ru}). 
Thus, reversing the sign of $\Lambda$ is similar to changing 
$\theta_{LD}\rightarrow\theta_{LU}$ and $\theta_{RD}\rightarrow\theta_{RU}$, or 
in other words, changing $y\rightarrow-y$. Hence, reflecting the whole system 
about its longitudinal axis, the particles chirality will be reversed, see Fig. 
\ref{fig18} which is the counterpart of Fig. \ref{fig14}(a), for $\Lambda=+1$. 
In this case, $A$ and $B$ edge states will have a positive 
momentum, while $C$ and $D$, a negative one. Recall Fig. \ref{fig15}(a). From a 
transport perspective, this 
is equivalent to reversing the 
conduction 
channels and, consequently, the Hall resistance and Chern numbers transform as 
$R_H\rightarrow-R_H$ and, respectively, $C^{(n)}\rightarrow-C^{(n)}$.

\section{Summary and Conclusions}\label{IIIII}

In this paper, we investigated the topological properties of a 2D SSH system 
from the perspective of FTIs. TRS breaking was 
achieved through circularly polarized light irradiation. Additionally, besides 
the topological phases governed by Chern numbers, the system may also undergo a 
 topological phase transition due to 
dimerization state switching. 
This transition is generally specific to SSH 
systems and is independent of light 
driving.

Our aim in this research was to fundamentally describe the interplay between 
TRS breaking and dimerization. We began 
our work by revisiting the well-known 1D SSH 
chain, highlighting its most important topological properties to serve as a 
benchmark for interpreting our results. Next, we introduced the 2D SSH lattice 
model, which allows for diagonal hopping between second-order neighbors. After 
formulating the 2D SSH Hamiltonian and utilizing Peierls substitution, we 
developed a model for interaction with circularly polarized light.

One significant finding was that TRS breaking, and implicitly, a topological 
phase transition, is not allowed in the absence of diagonal hopping. Moreover, 
this result was fully supported analytically.

Further, we performed a comprehensive numerical analysis to investigate the 
fundamental properties of the studied model. Initially, we examined the 
scenario without light, which served as our starting point.
We showed that the degeneracy points at $\Gamma$ and $M$ are 
protected by the invariance under the action of the  
anti-unitary operator $\Upsilon$ (\ref{upsilon}) [Eqs. (\ref{rcom}) and 
(\ref{rcom2})], and that the reflection 
symmetry 
(\ref{rcom3}) 
guarantees the flat band appearance along the $\Gamma$-$M$ direction.

 Employing the 
decimation method, we proved that the lattice could be interpreted as being 
formed of 'composite' atoms, in which previous atoms are arranged in groups of 
two. The newly formulated model proved to be invariant under the action of 
$\sigma_x$, allowing us to interpret the four atomic states ($A$-$D$ and 
$B$-$C$) as conserved spin states, effectively serving as an internal degree of 
freedom for the two "composite" atoms of the effective unit cell. Given that 
the spin states are related by time reversal operation, we inferred that TRS 
breaking occurs between $A$ and $D$ states and similarly between $B$ and $C$. 
Consequently, we anticipated that in a topological phase where chiral edge 
states arise, the mixing between $A(B)$ and $D(C)$ is topologically forbidden. 
In other words, one edge of the system 
will
host
$A$ and $B$ states, 
while the other, $C$ and $D$, respectively.

Next, we introduced light irradiation to break the TRS and computed the 
topological phase diagram based on Chern numbers. We have thus found an 
interesting property. For 
$\gamma_2<\gamma_1$, it is impossible to induce a phase in 
which all the three Chern numbers are non-zero and 
have interpreted this 
behavior in terms of the effective model of bipartite unit cell. By imposing 
$\gamma_2=\gamma_1$, we demonstrated that a gap is always reserved for 
a topological phase transition due to the
dimerization state switching, 
occurring at the non-bonding energy, as typically observed in SSH systems. 
Additionally, we highlighted the specific sublattice polarization of the 
topological bands in both infinite and ribbon configurations. Using the local 
density of states (LDOS) function, we confirmed the prediction that $A(B)$ and 
$D(C)$ states are separately localized at the system
edges, giving rise to 
conduction channels with opposite chirality in a topological phase. 
Furthermore, we discussed the implications in terms of quantum Hall transport, 
particularly the formation of $R_H$ plateaus.

Finally, we explored the case of 
$\gamma_2>\gamma_1$. Here, we demonstrated that the 
phase diagram becomes more intricate and provided examples of simultaneous 
non-zero Chern numbers. For such a topological phase, we presented the 
quasienergy dispersion, and in the quantum Hall transport, a new 
$R_H=+\frac{1}{2}$ plateau represents the signature of the 
new topological phase induced by the dimerization state 
switching. Moreover, we 
explored the scenario of light helicity reversal $\Lambda\rightarrow-\Lambda$, 
resulting in the switching of particle chirality.

The most significant findings from our research are that: (i) TRS breaking is 
not achievable in the absence of second-order neighbors (diagonal) 
hopping, and (ii) there 
exists an interplay between TRS breaking and 
dimerization, reflected in the values of the Chern 
numbers $C^{(n)}$.
\section{Acknowledgements}
We acknowledge financial support from the Core Program of the National 
Institute of
Materials Physics, granted by the Romanian MCID under Project No. 
PC2-PN23080202.
\appendix
\section{Landauer-B\"uttiker formalism}

If we consider a potential
$V_1-V_4$ applied between lead 1 and lead 4 and measure $V_2-V_3$ between lead 
2 
and lead 3,
the 
Hall resistance reads 
\begin{equation}
R_H=\frac{V_2-V_3}{I_1},
\end{equation} 
where $I_1$ is 
the current which flows at terminal 1. Leads 2 and 3 are probes, thus 
$I_2=I_3=0$. Setting $V_4=0$, the Ohm's law reads: 
\begin{gather}
G
\begin{pmatrix}
V_1\\
V_2\\
V_3\\
\end{pmatrix}
=
\begin{pmatrix}
I_1\\
0\\
0\\
\end{pmatrix};\label{ohm}\\
G=
\begin{pmatrix}
g_{12}+g_{13}+g_{14}&-g_{12}&-g_{13}\\
-g_{21}&g_{21}+g_{23}+g_{24}&-g_{23}\\
-g_{31}&-g_{32}&g_{31}+g_{32}+g_{34}\\\label{cond}
\end{pmatrix},
\end{gather}
where $G$ denotes the conductivity matrix with $g_{ij}$, the conductivity 
function 
between $i$ and $j$ leads. Eq. (\ref{ohm}) may be also translated as
\begin{equation}
\begin{pmatrix}
V_1\\
V_2\\
V_3\\
\end{pmatrix}
=R
\begin{pmatrix}
I_1\\
0\\
0\\
\end{pmatrix},\label{ohm2}\\
\end{equation}
where $R=G^{-1}$ represents the resistance matrix. Finally, computing the 
inverse of (\ref{cond}), the Hall resistance is defined as
\begin{equation}
R_H=\frac{g_{24}g_{31}-g_{21}g_{34}}{\det(G)}\left(\frac{h}{e^2}\right),\label{rh}
\end{equation}
where $h/e^2$ is the resistance quantum. 

Within the Landauer-B\"uttiker formalism, $g_{ij}$ is numerically computed 
using the Green's functions method \cite{datta}.

\end{document}